# Prediction of NMR Parameters and geometry in $^{133}$Cs-containing compounds using Density Functional Theory

N. Manukovsky[1], N. Vaisleib[1], M. Arbel-Haddad[2], A. Goldbourt[1,§]

[1]School of Chemistry, Tel Aviv University, Ramat Aviv 6997801, Tel Aviv, Israel
[2]Nuclear Research Center Negev, PO Box 9001, Beer Sheva 84901, Israel

[§]amirgo@tauex.tau.ac.il

## Abstract

The need to immobilize low-level nuclear waste, in particular $^{137}$Cs-bearing waste, has led to a growing interest in geopolymer-based waste matrices, in addition to optimization attempts of cement matrix compositions for this specific application. Although the overall phase composition and structure of these matrices are well characterized, the binding sites of Cs in these materials have not been clearly identified. Recent studies have suggested that combining the sensitivity of solid-state Nuclear Magnetic Resonance (SSNMR) to the local atomic structure with other structural techniques provides insights into the mode of Cs binding and release. Density Functional Theory (DFT) can provide the connection between spectroscopic parameters and geometric properties. However, the reliability of DFT results strongly relies on the choice of a suitable exchange-correlation functional, which for $^{133}$Cs, the NMR surrogate for such studies, is not well-established. In this work we benchmark various functionals against their performance in predicting the geometry of various simple Cs compounds, their NMR quadrupolar coupling constants, and their chemical shift values, while prioritizing the ability to incorporate dispersion interactions and maintaining low computational cost. We examined Cs salts, Cs oxides, perovskites, caged materials, a borate glass and a cesium fluoroscandate. While no single functional performs equally well for all parameters, the results show rev-vdW-DF2 and PBEsol+D3 to be leading candidates for these systems, in particular with respect to geometry and chemical shifts, which are of high importance for Cs-immobilization matrices.

## Introduction

Cesium is an alkali metal whose ionic form plays a key role in various materials and applications, including chemical catalysis[1], purification of biological macromolecules[2], and optoelectronic devices[3]. Radioisotopes of Cs, specifically $^{137}$Cs, a β-emitter with a half life time of ~30 yrs, are found in radioactive waste streams. The long term immobilization of Cs-bearing radioactive waste is a challenging task due to the high solubility of Cs ions in aqueous media. Whereas vitrification is the method of choice for high-level nuclear waste[4,5], cementitious matrices are usually used for low-level and intermediate-level waste immobilization due to their low cost and ease of preparation, combined with good mechanical properties and durability under irradiation. However, cesium is not efficiently immobilized by the alkaline environment afforded by cement. Geopolymers have been proposed as an alternative for Cs-bearing low-level nuclear waste due to their ability to specifically bind Cs ions[6]. Examples from recent years are the many studies regarding immobilization of incineration ashes from the Fukushima region, which have a high content of radioactive $^{134}$Cs and $^{137}$Cs following the nuclear accident in 2011. Alkaline activation of the ashes leads to the formation of geopolymeric wasteforms[7–9] consisting of various crystalline and amorphous phases. The ultimate Cs immobilizing matrix should be characterized by high Cs binding, and consequently minimal leaching to the environment under relevant environmental conditions. High-resolution structural characterization of Cs-bearing matrices, clarifying properties such as the binding site(s) geometry, the binding phase where several phases coexist, and the equilibrium distribution between competing sites, can contribute to the rational design of improved immobilization matrices.

Parameters provided by solid-state NMR (SSNMR), such as chemical shift, dipolar coupling, and quadrupolar coupling, are sensitive to the local geometry of the studied nuclei, including its binding atoms, bond lengths and angles, and its non-covalent interactions with the environment. Thus, they contain a wealth of local geometric and electronic structure information. However, only few examples show how $^{133}$Cs SSNMR can be utilized to characterize Cs binding sites in nuclear waste matrices[4,10–17].

In general, there is no simple, straightforward method to directly infer structural properties from the NMR parameters unless the dipolar interaction, which is proportional to the inverse cube of the distance, is directly measured. First-principles calculations are known to bridge the gap between spectroscopy and structure, thus allowing to utilize the information content of SSNMR to its fullest. Density functional theory (DFT) has been in vast use in the physics and chemistry communities for decades due to its ability to predict diverse properties such as ground state structure, excitation energies, spectroscopic properties, reaction energies and more, for molecules and solids, with high accuracy and at a relatively modest computational cost. Specifically, the use of DFT calculations side by side with NMR measurements has become widespread since the introduction of the GIPAW (Gauge Including Projector Augmented Wave) method[18–20], which provides the ability to compute NMR tensors in solid systems with all-electron accuracy. Although DFT has been helpful in the structural characterization of zeolites[21,22], $^{133}$Cs has not been extensively studied using DFT-GIPAW. Relevant examples of the combination of $^{133}$Cs SSNMR and DFT are the study of Cs adsorption on Montmorillonite Clay[12], and their use to study Cs-containing perovskites[23] and cryptands[24].

Prior to the use of DFT-GIPAW for probing Cs binding in complex systems such as cements and geopolymers, which are often a conglomerate of several crystalline and non-crystalline phases, it is instructive to screen the suitability of different DFT functionals to correctly predict $^{133}$Cs NMR parameters using simpler, well characterized Cs-containing systems. Cs-halides (CsF, CsCl, CsBr, and CsI) are the obvious place to start. Experimentally, it is well known that under ambient temperature and pressure, CsF crystallizes in the B1 phase (Figure 1a), forming the so called NaCl or rock-salt structure, as most alkali halides do. However, CsCl, CsBr, and CsI crystallize in the B2 phase (Figure 1b), being the only alkali halides known to crystallize in the so-called CsCl structure under these conditions[25] (the discussion omits Fr and At, about which no data is available)[26]. The reason for this difference is dispersion interactions which favor the phase with the shorter distances between ions of the same kind, that is, the CsCl phase. These interactions depend on the ion polarizability, and are therefore more substantial in the mentioned systems consisting of heavy ions[27,28]. Van der Waals forces make only a minor contribution to the total lattice energy of an ionic lattice (1-5% of the lattice energy), but they are the factor determining the phase in this case[28]. Zhang et al.[29] have shown that despite the simplicity of cesium halide structures, the popular Perdew-Burke-Ernzerhof (PBE)[30] functional, commonly used for the prediction of NMR parameters for $^{133}$Cs[12,23] and other nuclei[31–34], fails to predict the correct phase for CsCl. The same also occurs when using other GGA-type functionals such as the solid-state optimized PBEsol[29], PW91 and revPBE[35], as well as for various meta-GGA type functionals (MS0, MS1, MS2, TPSS and rTPSS)[35]. A plausible explanation is that pure DFT functionals do not include terms describing dispersion forces[29,36]. Pure DFT functionals are semilocal, meaning that the energy density at any point depends only on properties at that point. Consequently, these functionals are by definition unable to model dispersion, which is an interaction between instantaneous dipoles at two different sites. Indeed, the correct CsCl structure is obtained using calculation schemes incorporating dispersion[29,37].

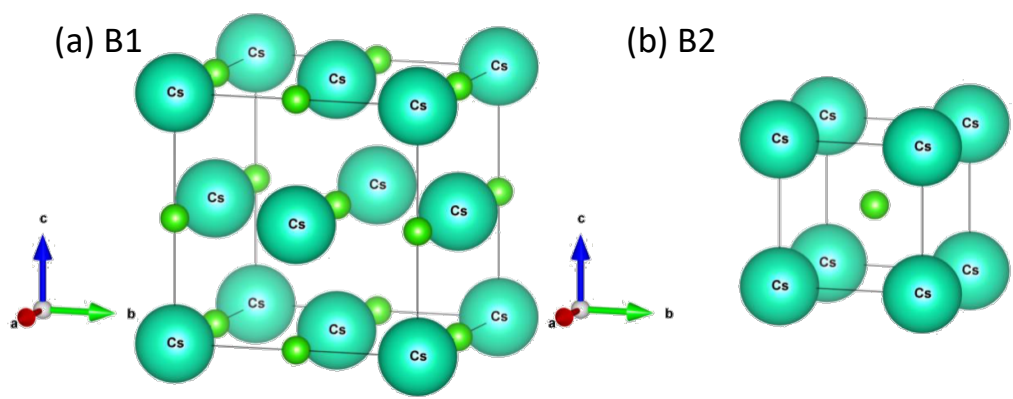

Figure 1: (a) B1 (NaCl) phase vs (b) B2 (CsCl) phase unit cells[25]. The large spheres represent the metal, whereas the small ones represent the halide. At room temperature, only CsF adopts the B1 phase.

Dispersive interactions can be introduced into DFT by various approaches[36]. A common approach is the addition of asymptotic dispersion corrections to the energy obtained by a standard DFT calculation, in the manner of an interatomic potential[38–45]. Notable examples of dispersion corrections are the Grimme schemes DFT+D[38], DFT+D2[39] and DFT+D3[46]; Tkatchenko-Scheffler (TS) dispersion correction[40] and the subsequent Many-body dispersion (MBD) correction[47], and Exchange-hole dipole-moment model (XDM)[48–50]. Asymptotic dispersion corrections can attain good accuracy at a relatively low computational cost, and can be applied in conjunction with various, well-tested functionals. Another approach is the development of explicitly non-local density functionals[51–58], whereby the energy depends on the density and its gradient at two points in space simultaneously. Notable examples of non-local functionals are vdW-DF[53], vdW-DF2[56], vdW-DF3[59], and rVV10[54,57]. Additional methods include dispersion-correcting potentials[60–63], and parameterized exchange-correlation functionals[64–66]. Finally, meta-GGA functionals[67,68] can incorporate short-range, but not long-range, dispersion interactions through their dependence on the kinetic energy.

For Cs halides, forcing the right phase upon the popular PBE functional yields a good prediction of the chemical shift[69], but does not solve the problem of the accurate geometry prediction. It is currently unknown which functional can correctly capture both the geometry and the NMR parameters of Cs-compounds, ranging from Cs halides to Cs-bearing Zeolites.

This work aims to identify the most suitable DFT scheme for the prediction of the correct geometry and NMR parameters of various Cs-containing systems. First, we screen various functionals for their ability to predict the correct phase in Cs halides. We then compare their performance in predicting the crystal unit cell volume, chemical shift and quadrupolar coupling constants in Cs halides, Cs oxyanion compounds, and Cs-containing perovskites. Keeping in mind our interest in more computationally-demanding Cs-bearing materials, we have restricted this study to schemes of a reasonable computational cost. Differences in the performance and accuracy of various functionals, as well as their performance compared with experimental results available in the literature, are discussed, and recommendations for functional selection for the prediction of $^{133}$Cs NMR parameters are proposed.

## Computational Methods

DFT calculations were carried out using the Quantum Espresso 7.1[70,71] software package, which uses the gauge-including projector augmented waves (GIPAW) method for computing magnetic resonance properties of crystalline structures. We used scalar relativistic PAW PBE or PBEsol pseudopotentials[72], downloaded from the Quantum Espresso website. Spin-orbit coupling (SOC) effects on the geometry were found to be sufficiently small to be neglected at this stage and were not further considered (see SI Table S13). The plane wave cutoff energies for eighteen cesium-containing systems, determined by convergence tests, are listed in Table S1 in the SI. The kinetic energy cutoff was four times that of the wave function. Monkhorst-Pack grids were used to sample the Brillouin zone (see Table S1 for their dimensions), and the initial structures were taken to be the reported crystal structures (Table S2).

Various functionals and dispersion corrections were used for the calculations of geometry and NMR parameters. PBE[30], known to fail in predicting the correct phase in cesium halides, is used as a reference, being a very common functional. Other GGA-type functionals examined here are revPBE[73], a one-parameter modification of PBE; its further modification rPBE[74], which performs well for the calculation of NMR parameters in solids containing elements of the third and fourth periods[75,76]; PBEsol[77], tailored to solids; and B86bPBE[30,78], the recommended functional to be paired with XDM for solid state calculations[44]. These five semilocal functionals are used as base functionals in conjunction with three dispersion corrections – DFT-D2[39], DFT-D3[46], and XDM[50,79]. PBE+MBD was not used as it is known to fail to converge for many cesium halides, and neither was PBE+TS which is known to yield the correct phase but provide incorrect lattice constants[80]. Non-local functionals considered are rVV10[58], vdW-DF1[53], vdW-DF2[56], vdW-DF3[59], vdW-DF-C6[81], rev-vdW-DF2 (also called vdW-DF2-B86R)[82].

Asymptotic dispersion corrections (such as D2, D3, and XDM) are implemented post-self-consistently, meaning they do not affect the electron density. Therefore, for a given geometric structure, the NMR properties, which strongly depend on the electronic structure, will be unaffected by these corrections. In contrast, nonlocal functionals are formulated self-consistently, directly influencing the electron density. However, all dispersion corrections of any type can indirectly affect any molecular property via their effect on the geometric structure. For this reason, we found it instructive to compare the NMR parameters obtained by different functionals using either the experimental (crystallographic) geometry or the DFT-optimized geometry, allowing us to distinguish between geometry-mediated effects and electronic-structure-mediated effects. In the cases where structural relaxation was implemented, both the lattice constants and the atomic positions were optimized using the BFGS algorithm, until the residual forces acting on each atom were below 1.0E-03 Ry/au.

## Results and Discussions

### Cesium halide polymorphs

Pure DFT functionals are known to fail in the prediction of the ground state structure of heavy Cs halides, owing to the absence of dispersion in the energy calculation. Therefore, the first criterion for DFT scheme screening was the ability to correctly predict the ground state phase, which is B1 for CsF and B2 for heavier Cs halides. Unfortunately, the comparison to experiment is only qualitative, as the experimental values of the energy gap between Cs halides polymorphs are unavailable.

The results confirm that semilocal functionals erroneously assign the B1 phase to all Cs halides (Figure 2a). Interestingly, the slope of the energy gap curve is the same for all five GGA-type functionals tested here, with the exception of PBEsol for the CsF-CsCl segment. The addition of an asymptotic dispersion correction leads to the correct phase prediction in most, yet not all, combinations of base functionals (PBE, rPBE, revPBE, PBEsol, B86bPBE) and dispersion corrections (D2, D3, and XDM) tested here. The D2 correction (Figure 2b) using the default parameters leads to a non-physical trend where the energy gap is non-monotonous versus the halide size and was therefore not further used. Using Zhang's parameterization for Cs ($C_6^{Cs} = 57.74 \frac{J \cdot nm^6}{mol}, R_0^{Cs} = 1.776$ Å)[29] rectifies this. Using Schurko's D2* reparameterization of the damping function[83] (d=3.25 instead of the default 20), originally implemented using rPBE, on its own leads to a nearly flat curve and a wrong phase in CsF. Using Zhang's parameters together with D2* does not settle the phase issue and does not reproduce the trend with respect to the halide size. We note that D2* was parametrized using the EFG tensors of three elements, and should in principle be transferrable to all elements. However, it was designed to provide a good fit of the EFG, not necessarily of the phase stability, and was trained on small organic molecules, very different from the inorganic salt structure studied here.

The D3 and XDM corrections (Figure 2c) lead in most cases tested here to the correct phase, although the trend is violated for PBE+D3, PBEsol+D3, and B86bPBE+XDM (the curves go slightly up from CsBr to CsI). A similar behavior was described for PBE+D3 with Li halides[84]. The energetic gap between the phases is larger for PBE (or PBEsol)+D2 than for PBE (or PBEsol)+D3. This can be explained by the fact that D2 describes dispersion using only two-body attractive interactions, whereas D3 also incorporates local environment effects and three-body repulsive interactions[85].

All vdW-DF functionals tested (Figure 2d) yield curves of the same slope. Most of them yield the correct phase, as does rVV10. The hybrid functionals tested here do not capture the correct phase.

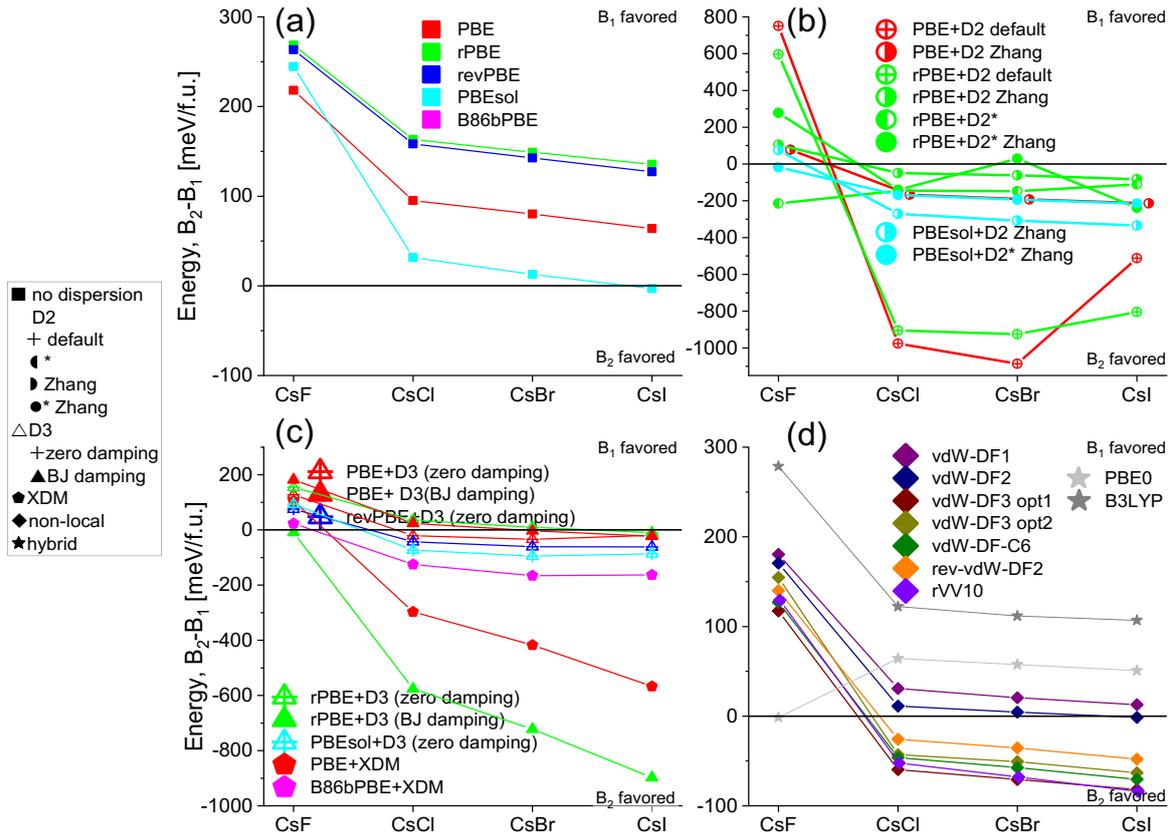

*Figure 2: Relative stability of Cs halide phases for various functionals: (a) semilocal functionals (■); (b) semilocal functionals with DFT-D2 dispersion correction (○; the symbol filling indicates the parametrization used); (c) semilocal functionals with DFT-D3 dispersion correction (△, the filling indicates the type of damping function used) or XDM (⬠) dispersion correction; (d) non-local (◆) and hybrid (★) functionals.*

The reliable structure prediction of Cs halides is challenging, since the energy gap between competing phases, caused by the typically weak dispersion forces[28], is small, and is therefore very sensitive to errors arising in various approximations of the exchange-correlation energy. The energy gaps predicted here are similar to those calculated by Pyper (0.078 eV/f.u.)[86], Aguado (0.14 eV/f.u.)[87], and Zhang (0.16 ev/f.u.)[29]. As shown by others[29], GGA and hybrid functionals fail to predict the correct phase of Cs halides. We therefore chose to proceed with non-local and dispersion-corrected semilocal functionals.

### Unit cell volume

In the next step of this study the ability of different functionals to predict the unit cell volume of Cs compounds was tested. The panel of Cs-compounds was expanded to include additional systems, namely four Cs oxyanion salts and three $CsGeX_3$ perovskites in addition to the four Cs halides studied thus far. Figure 3 features unit-cell volumes obtained from X-ray diffraction (literature values) and calculated unit cell volumes obtained using various DFT schemes, along with the mean error (ME) for each functional. We note that the DFT results refer to 0 K, while the experimental structures were determined at or around room temperature.

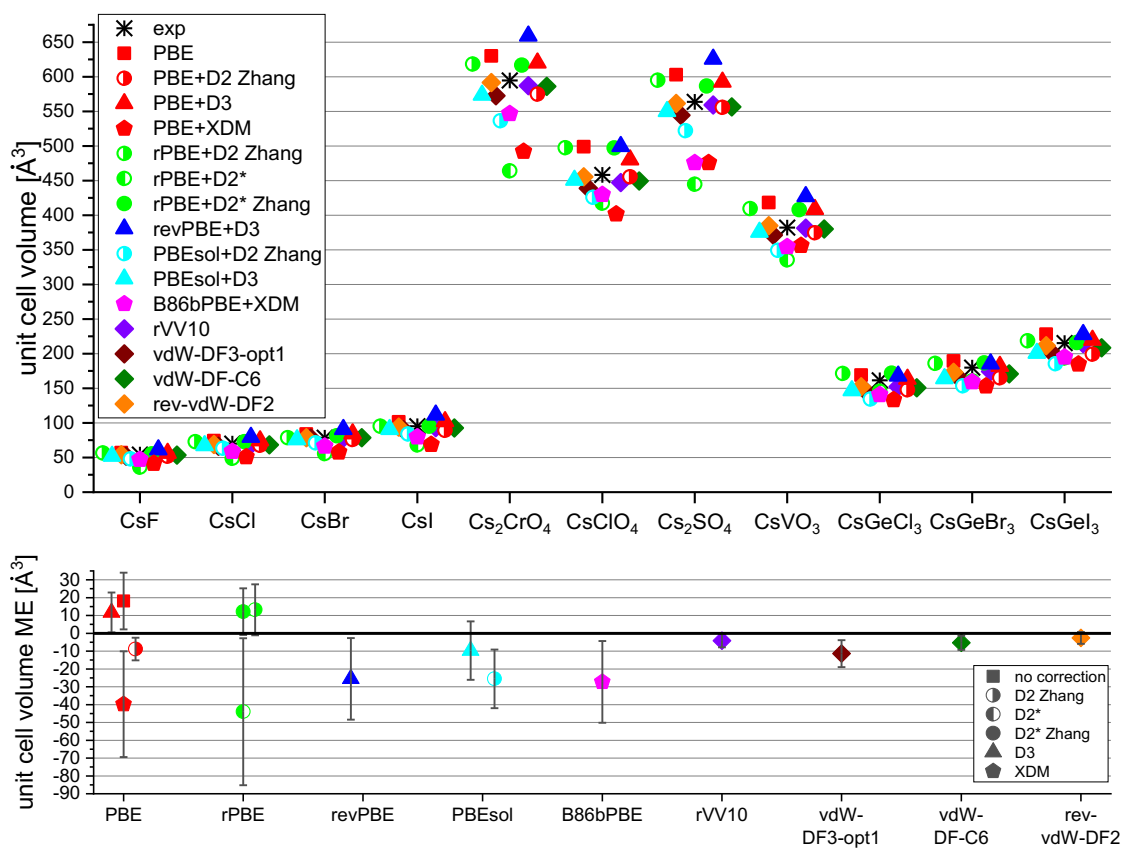

*Figure 3: Unit-cell volume of various Cs-containing compounds as found experimentally by X-ray crystallography (asterisk) or by calculation using various DFT schemes. Some of the points are slightly horizontally shifted to avoid overlap. See Table S2 for references of the experimental data.* The bottom figures show the Mean Error (ME) of each functional, the error bars representing the standard deviation (1 SD above and below the mean). The mean relative error (MRE) and the coordinates root-mean-square-error (RMSE) are found in Figure S1.

Most schemes give a reasonable prediction of the unit cell volume, and correctly capture the trend among different compounds. Quantitatively, however, there are differences between various schemes. As expected, PBE overestimates the unit-cell volume. Adding a dispersion correction to PBE results, as expected, in a contraction of the calculated unit cell, to an extent that depends on the type of dispersion correction used. D2 is superior to XDM and D3 in this context, with XDM leading to an underestimation of the cell volume. The functionals revPBE+D3 and B86bPBE+XDM are outperformed even by uncorrected PBE. This is in contrast to the good performance of B86bPBE-XDM in predicting the lattice constant of the entire series of alkali halides[88]. This may be related to the fact that unlike the former study[88], the current work includes compounds containing polyatomic anions. Good performance is also displayed by the non-local functionals, as well as by rPBE+D2* Zhang parametrization and PBEsol+D3. The tested non-local functionals are the best performers in this task, as manifested by both their small Mean Error and their small standard deviation values.

### $^{133}$Cs NMR parameters: The quadrupolar coupling constant

The calculation of the $^{133}$Cs quadrupolar coupling constant $C_q$, using GIPAW and the same DFT schemes, is shown in Figure 4. A great difference is noticeable between different compounds: some are very sensitive to the calculation scheme (e.g. $CsClO_4$, $CsVO_3$), whereas some are not (e.g. $CsGeX_3$, $Cs_2CrO_4$ site 2, $Cs_2SO_4$ site 1). This observation highlights the importance of screening a large sample of materials before choosing which calculation scheme to use.

PBE, though not accounting for dispersion and not providing an accurate unit cell volume (Figure 3), provides a very good prediction of the Cq parameter, slightly outperformed only by PBE+D3 and rPBE+D2* Zhang. Adding D2 or XDM to PBE leads to an overestimation of |Cq|, although PBE+D2 yields good geometry (Figure 3). The other combinations of a GGA functional with an asymptotic dispersion correction are not better in terms of absolute error, but adding Grimme's D3 correction improves the relative error. All four non-local functionals yield similar results and are all inferior to PBE and to PBE+D3, although their unit cell volume prediction was better (Figure 3).

Interestingly, when using the experimental geometry without geometry optimization (empty squares), all schemes tested here provide nearly identical results. It is not surprising that the semilocal functionals provide results similar to each other, as do the non-local functionals. However, it is not obvious that these two different families of functionals provide results similar to each other, despite the fact that dispersion is inherent in the non-local functionals, while this is not the case in the corrected ones. We therefore conclude that the main effect determining the accuracy of Cq here is the geometric structure, not the difference in the functional itself. This is in agreement with other studies[89]. However, remarkably, the functionals excelling at geometry prediction (e.g. rVV10, rev-vdW-DF2) do not excel at Cq prediction, showing that the structure most similar to the experimental one is not necessarily the one leading to the best Cq, at least for the type of Cs compounds studied here. Moreover, geometry optimization often results in a higher deviation in the Cq predicted value from that of the experimental value, with the exception being uncorrected PBE, PBE+D3, and rPBE+D2* Zhang parameterization. We note that the small sample of materials used here, owing to the scarcity of experimental $^{133}$Cs NMR results in the literature, results in large standard deviation values. Still, a fairly linear correlation is maintained between calculated and experimental Cq values (**Error! Reference source not found.** and **Error! Reference source not found.**).

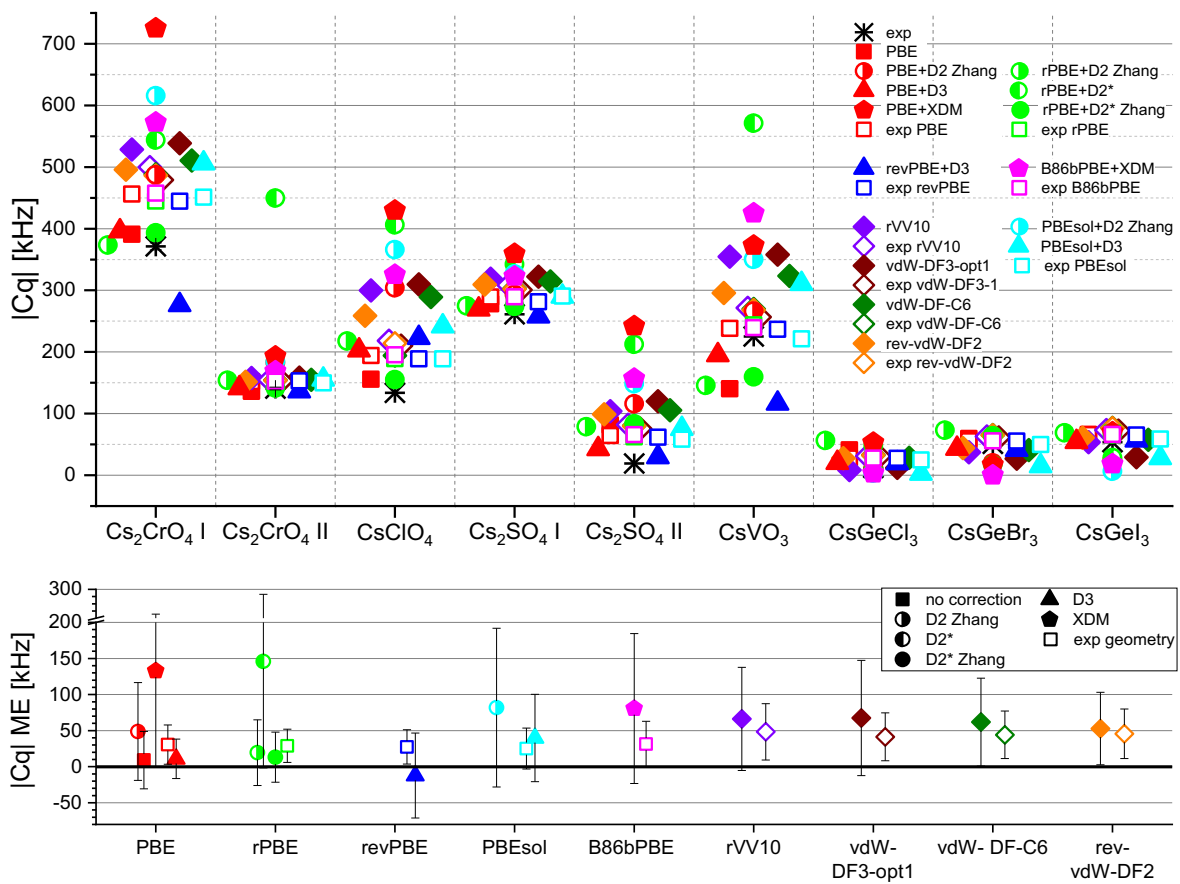

*Figure 4: Quadrupolar coupling constant Cq of various Cs compounds. The black asterisks correspond to the experimental values (Table S3), whereas the colored symbols correspond to computational results (Tables S4-S6). Solid symbols represent calculations where the geometry was optimized, whereas open symbols correspond to calculations employing the experimental (XRD) geometry.* The bottom plot shows the mean error in Cq. Additional error metrics are given in Figure S1. The error bars represent the standard deviation.

An important aspect in the analysis of $^{133}$Cs, is that in many materials of interest the value of its quadrupolar coupling constant is small, and its experimental measurement is therefore difficult, also due to its high spin (S=7/2). The chemical shift, which has a relatively broad ppm range, becomes the more informative $^{133}$Cs NMR observable. We therefore turn to calculations of $^{133}$Cs shifts.

### $^{133}$Cs NMR parameters: Chemical shifts

We have used DFT calculations to predict the $^{133}$Cs chemical shielding tensor. To convert the absolute shielding to the chemical shift, we plotted in Figure 5a the calculated isotropic shieldings versus the experimental isotropic shifts. Experimental values have been obtained from various literature sources, as shown and referenced in Table S8, as well as experiments performed in our lab. A literature review revealed that for the simple Cs salts we analyzed, different reference compounds have been used and occasionally spectral referencing was not clearly stated. $^{133}$Cs chemical shifts are known to be concentration dependent[90], and the use of CsCl solutions as an external reference must be done with care, in particular when using hygroscopic compounds such as CsNO$_3$. The commonly cited manuscript by Haase[91] contains data from experiments performed on static samples and at a low field (2.1T), and reports that a solution of 0.5m CsCl resonates at +6.1 ppm with respect to infinite dilution, and that solid CsCl is at 228.1 ppm. A later study by Mooibroek[92] reports solid CsCl to be +223.2 ppm with respect to 0.5m CsCl, and this value is more commonly used in the literature, although various studies also use a 0.5M or 1M CsCl solution as the reference value of 0 ppm. Our experimental data, measured under 5 kHz magic-angle spinning at a field of 14.1T, shown in Figure S4, allowed us to reconcile the seemingly contradicting literature results, setting a uniform reference scale. We used solid CsCl as a reference at +223.2 ppm, obtained that a 1M CsCl solution has a chemical shift of 5.0 ppm, and from several solutions of CsCl the infinite dilution chemical shift value is projected to -6.1 ppm, in agreement with Haase and the correction by Mooibroek for solid samples. All experimental values obtained from the literature were therefore re-referenced using the corrected scale. To demonstrate the usefulness of using solid CsCl as a reference, we also show in Figure S4 that its chemical shift is independent of temperature, whereas for 0.1M $CsNO_3$ in $D_2O$ (IUPAC recommendation), a shift of 1.9ppm is observed over 15°C.

The correlations between the re-referenced experimental chemical shifts and calculated chemical shielding, presented in Figure 5a, reveal that most calculation schemes are well fitted by linear curves, with the exception being rPBE+D2*. However, the slopes differ from the ideal value of -1 (Figure 5). The greatest deviations from the ideal slope are those of XDM corrected functionals and of rPBE+D2*, the latter being greatly improved by adding Zhang's parametrization along with that of D2*. The slopes closest to -1 are those of revPBE+D3 (-0.90±0.03) and PBE+D3 (-1.18±0.04). Most other functionals feature slopes in the range [-1.45, -1.2], the differences between most functionals being insignificant, even when comparing, for example, between uncorrected semilocal functionals using the experimental geometry and non-local ones including geometry optimization.

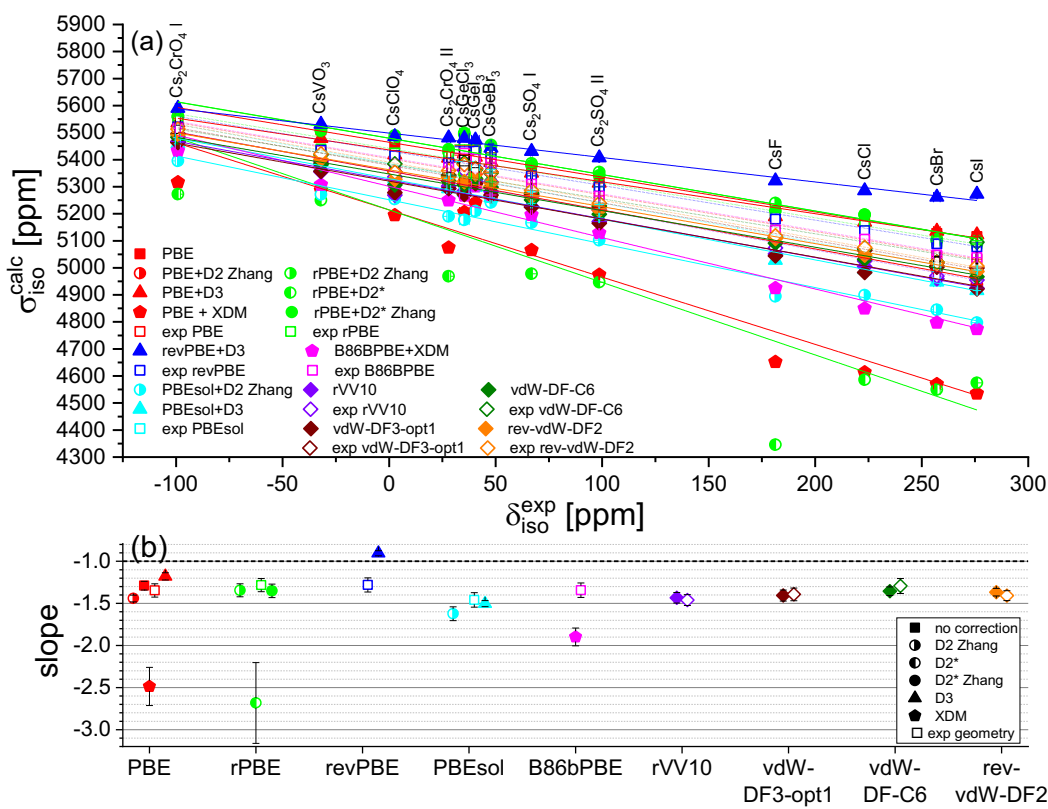

*Figure 5: (a) Linearization plots of the isotropic shielding of various schemes and (b) their corresponding slopes.* The error bars of the slopes represent standard deviations. $\delta_{iso}$ are averaged experimental results collected from the literature (re-referenced) and from our own measurements (Table S8 and Figure S4).

We used the fitted correlations to convert the calculated chemical shieldings to shifts. The results, plotted in Figure 6, show that all functionals capture the general trend in the chemical shifts of different systems. However, quantitative accuracy differs between different schemes. When using the experimental geometry, the shift is nearly independent of functional, as reported in other studies[93], and as we observed in the prediction of Cq. Geometry optimization improves the accuracy for most schemes, in contrast to the case of Cq calculation. While for experimental geometry the Mean Absolute Error (MAE) of $\delta_{iso}$ values is mostly between 15-20 ppm (empty symbols), the best accuracy after optimization is achieved by PBEsol+D3 (MAE=6.2 ppm), rev-vdW-DF2 (8.9), PBE+D2 (9.7), revPBE+D3 (10.4), PBE+D3 (10.5), rVV10 (10.8), and vdW-DF-C6 (10.9). These MAE's amount to 3% of the total shift span (372 ppm). Interestingly, it has been shown that PBE+D3 does not perform well in the calculation of $^{13}$C chemical shift in zeolites[93], whereas PBEsol+D3 is usually not used for chemical shift calculations.

Again, the limited availability of experimental $^{133}$Cs NMR results manifests itself in broad error margins, precluding a definite choice of the best-performing functional. However, the functionals displaying a small Mean Absolute Error also tend to display a lower standard deviation compared to the others, supporting their identification as good performers. Another comment regarding the variability of the average error, is that this variability is small for the unit cell volume (Figure 3), but large for the NMR parameters (Figure 4, Figure 6), although the sample size is nearly the same for both (not identically the same due to two of the systems containing two Cs-sites each). This suggests that for many functionals, the performance in geometry optimization is quite the same for different materials, but the performance of GIPAW for NMR parameters calculation is more sensitive to the system.

To distinguish between a systematic tendency rather than a random error in the estimation of the chemical shift, we also show the Mean Signed Error in Figure S1. For all tested functionals, the MSE is practically zero, indicating a random error.

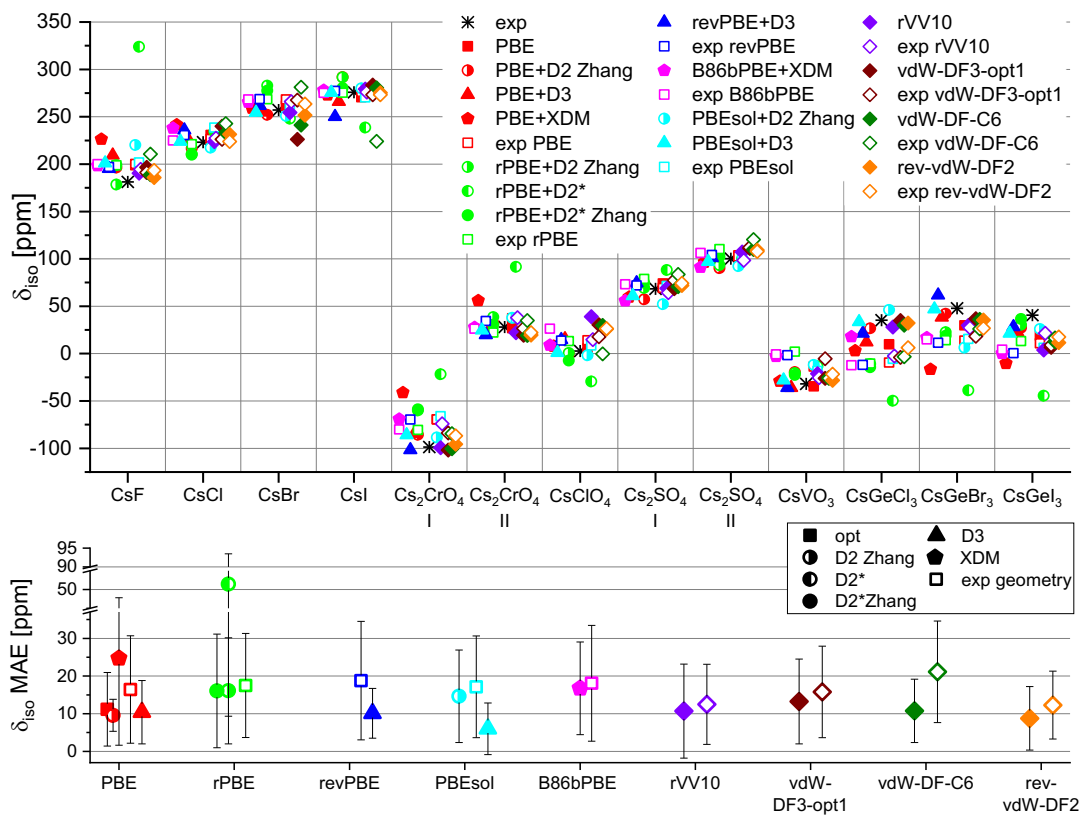

*Figure 6: Computational and experimental values of the isotropic chemical shifts for various Cs compounds.* Experimental values (Table S8) were partly measured by us (spectra shown on Figure S4) and partly by others. The bottom plot shows the mean absolute error in $\delta_{iso}$. *Error bars represent Standard Deviation.* An additional error metric is found in Figure S1. The calculated shifts, after linearization, are explicitly given in Tables S9-S12.

When zooming in on a more subtle trend, that of the isotropic chemical shift versus the halide size (Figure 7), and observing the apparently best functionals for the chemical shift prediction, it can be seen that revPBE+D3 fails to predict the trend in CsI, whereas rev-vdW-DF2 and PBEsol+D3, which were successful for the Cs halides, fail in CsGeX$_3$ (note that the parabolic shape, where the maximum is at Br, is kept). Only PBE+D3 qualitatively captures the halide trends in both series.

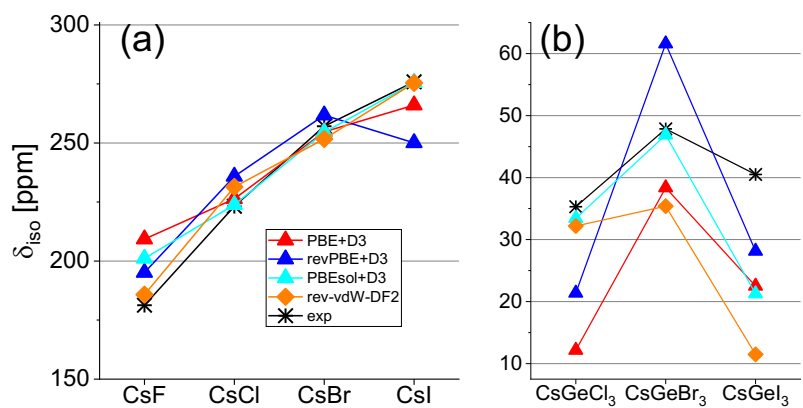

*Figure 7: Isotropic chemical shifts for (a) Cs halides and (b) Perovskites (Cesium Germanium halides).*

Chemical shift anisotropy (CSA) values have also been calculated for the various materials, and are shown in Figure S5 in the SI.

## Overall performance of functionals – Cs halides, oxyanions, perovskites

The mean errors of all functionals (including geometry optimization) for the unit cell volume, isotropic shift, and quadrupolar coupling constant are summarized in Figure 8.

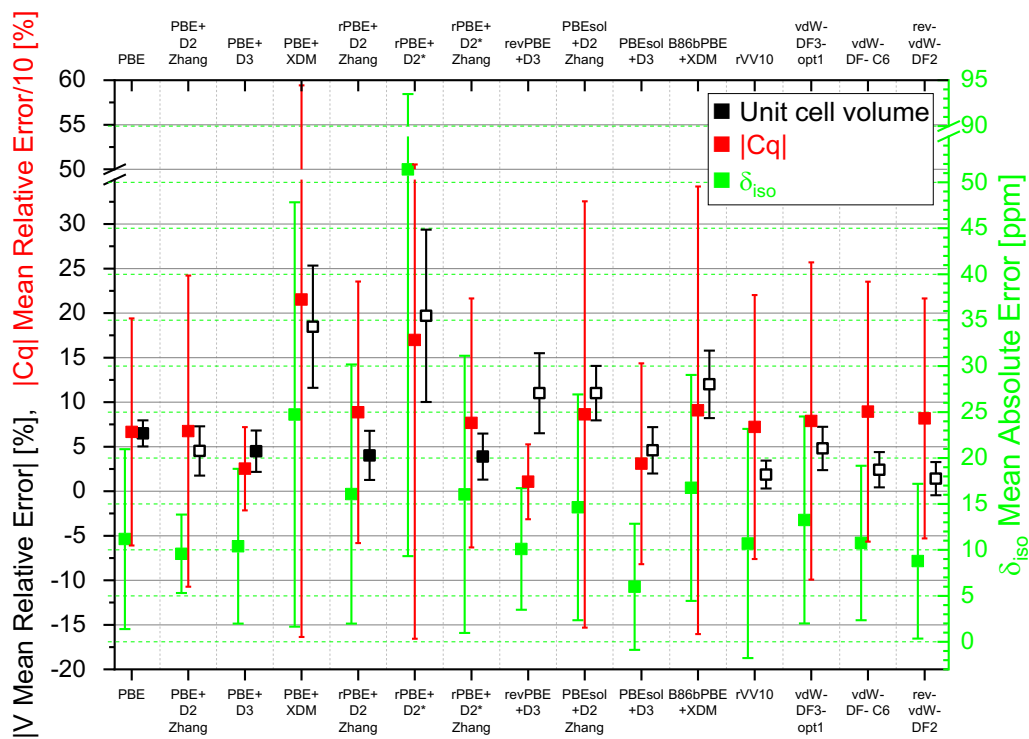

*Figure 8: MRE of the unit cell volumes (black, empty squares are for negative values), MAE of the isotropic chemical shifts (green), and MRE of the quadrupolar coupling constants (red) for all schemes. Symbols represent mean values; error bars represent Standard Deviation.*

PBE characteristically overestimates the unit cell volume, while providing good (yet not the best) NMR parameters. Adding the D2 correction leads to an underestimation of the cell volume while marginally improving the accuracy of the chemical shift. Adding the D3 correction to PBE leads to very good overall performance, among the best in each of the three observables. XDM correction to PBE or B86bPBE, leads to a major underestimation of the cell volume, as well as to poor chemical shift prediction. rPBE+D2, using Zhang's parametrization for Cs, provides good accuracy in the geometry, but not in the NMR parameters. Schurko's D2* parametrization has a marginal effect only. Interestingly, PBEsol+D2 fails in the geometry, but performs well for the NMR parameters (but with larger than typical standard deviations). Likewise, revPBE+D3 fails in the geometry, but surprisingly excels in Cq prediction, while performing well for the shift. PBEsol+D3 performs very well, among the best, for the volume and the quadrupolar coupling, and excels in chemical shift. Its overall performance is among the best.

rVV10, vdW-DF-C6 and rev-vdW-DF2 excel at the unit cell prediction, while performing among the best in the chemical shift, and provide reasonably good quadrupolar coupling constants. vdW-DF3-opt1 is slightly inferior to them.

To sum up, this panel shows PBE+D3, PBEsol+D3, rVV10, vdW-DF-C6 and rev-vdW-DF2 to display good overall performance in predicting the geometry and NMR parameters of Cs salts, oxyanions and perovskites.

### Extension to additional systems

To examine whether the results obtained using Cs salts, oxyanions and perovskites are transferrable to different Cs-containing systems, we applied the best performing functionals to very different systems, the structures of which are shown in Figure 9[94]. These systems differ from those in the set used in the previous section in several ways. $CsCd(SCN)_3$ (Figure 9a) is again a perovskite, but it contains a heavier element, Cd; in $CsBPh_4$ (9b) Cs interact with induced dipoles in the aromatic rings, unlike the ionic systems studied above; in [2.2.2] cryptand (9c) Cs interacts with electronegative atoms (O, N) in the altogether neutral complex via ion-dipole interactions, forming a stable coordination complex encapsulating the metal in a spheroidal cavity, somewhat analogously to Cs binding to zeolites, although with shorter Cs-O distances; $CsB_3O_5$ (9d) is a glassy system; in $CsSc_3F_{10}$ (9e) the coordination number of Cs is uncharacteristically high (18); $CsPbI_3$ (9f-g) is challenging due to the presence of the heavy Pb atom, in addition to Cs. Hence, we also compared in Figure S6 our calculations to those obtained including SOC showing that improvement is system dependent[95].

Here again we examine the prediction of unit cell volume, the quadrupolar coupling constant, and the isotropic chemical shift. The results are shown in Figure 10.

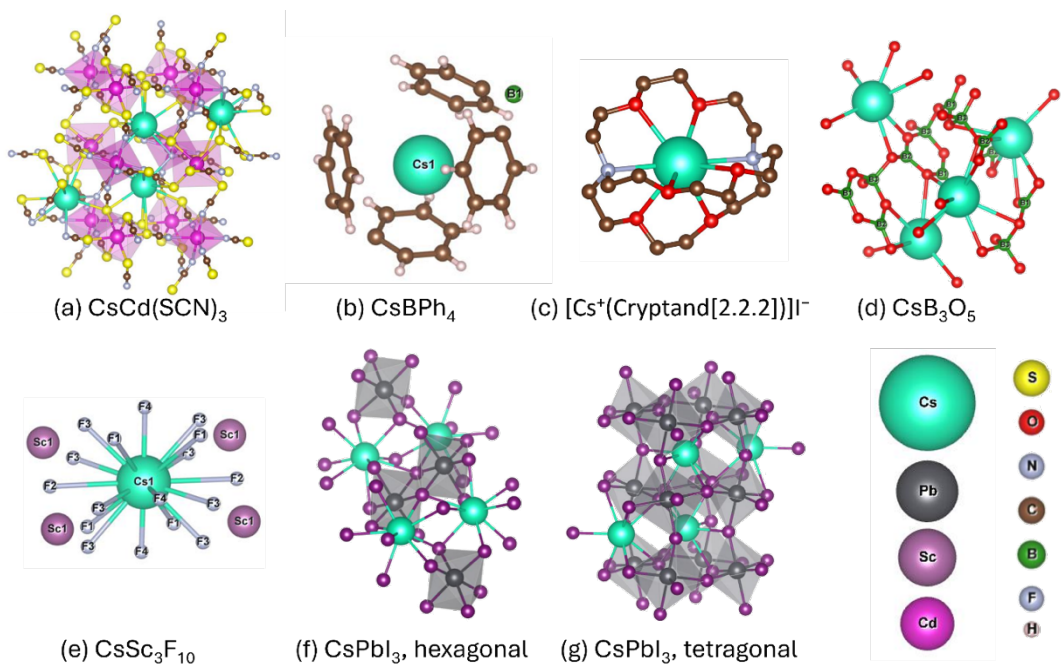

Figure 9: Structures of (a) $CsCd(SCN)_3$[96], (b) $CsBPh_4$[97], (c) [Cs+(Cryptand[2.2.2])]I−[94], (d) $CsB_3O_5$[98], (e) $CsSc_3F_{10}$[99], (g) hexagonal $CsPbI_3$[100], (h) tetragonal $CsPbI_3$[101].

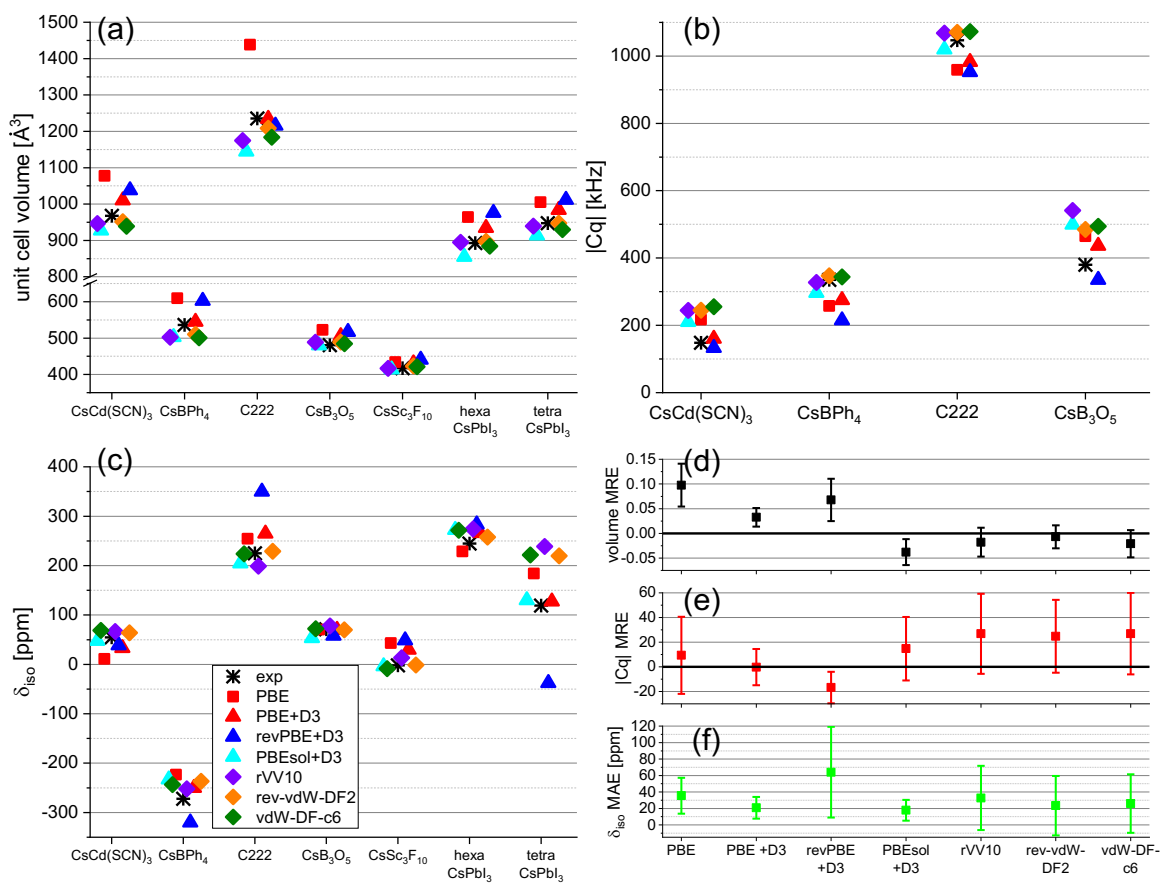

*Figure 10: (a) Unit cell volume, (b) Cq, (c) $\delta_{iso}$ and (d,e,f) error analysis. Symbols represent mean values; error bars represent Standard Deviation. See SI Table S8 for references of experimental values.*

As shown in Figure 10, here again the PBE functional overestimates the unit cell volume, while performing reasonably well (but not the best) in Cq. Its chemical shift performance is similar to that of most other functionals considered at this stage. PBE+D3 excels in the geometry, while being on-par with others in chemical shift, in line with the previous results (Figure 4, Figure 6). RevPBE+D3 shows the largest error in chemical shift, despite a relatively good geometry prediction, unlike the previous results. PBEsol+D3, which was the best performer in chemical shift for the simple systems, followed closely by rev-vdW-DF2, is again comparable with the latter in all three observables. All non-local functionals tested here excel at geometry prediction. Rev-vdW-DF2[82] has previously been shown to successfully predict the structural, mechanical, cohesive and vibrational properties of both weakly and strongly bound solids[102], as well as the interlayer binding energy of multilayered solids[103] and the adsorption energy of adsorbed molecules[104,105]. It is however not yet widely used for chemical shift prediction[106], and we are unaware of Cq calculations using this functional. We are also unaware of studies using PBEsol+D3 for the calculation of NMR parameters.

To conclude, switching from small systems to more complicated and diverse compounds, PBEsol+D3 and rev-vdW-DF2 maintain their overall good performance and relative advantage. While we cannot determine if this conclusion holds for other materials due to the relatively small (20) set of distinct Cs sites studied here, these functionals appear to be a promising choice for the determination of Cs geometry and NMR parameters, in particular chemical shifts that are more useful for distinguishing between the chemical environments of Cs.

## Summary & Conclusions

A series of DFT functionals were screened for their performance and accuracy in the prediction of the geometry, quadrupolar coupling constants, and chemical shifts of cesium-133 in Cs-based halides, oxyanions and perovskites. The functionals chosen were those considering dispersive interactions, in order to faithfully predict the phase of Cs halides, while keeping a low computational cost. The best performing functionals, chosen mainly based on their performance with respect to geometry and isotropic chemical shift predictions, were then tested on seven additional materials, which exhibit chemical environments, bonding and coordination that are different than the above-mentioned inorganic compounds, including interactions of the Cs cation with polar groups resembling to some extent Cs caged in zeolites. Overall, we examined 20 different Cs sites from 18 different compounds. While no single functional performs best for all parameters (geometry, chemical shift, quadrupolar coupling), the results suggest that two of the functionals tested, rev-vdW-DF2 and PBEsol+D3, maintained their advantage in predicting both geometry and isotropic shifts in these seven additional diverse materials.

Further work should be dedicated to the implementation of these functionals to the more challenging systems such as zeolites or geopolymers, used for the sequestration and immobilization of radioactive Cs species. More advanced (and computationally more expensive) meta-GGA and various hybrid functionals, possibly in combination with dispersion corrections may also be considered for these structurally complex systems.

As this work was the first stage of functional screening, it did not incorporate SOC, whose computational cost would have limited the number of tested systems and functionals. Further studies will have to consider the contribution of SOC to the accuracy and precision in predicting geometry and NMR parameters.

## Acknowledgements

This research was partially funded by the Pazy Foundation grant ID 701- 2025. We thank Orr Simon Lusky for preparing the $CsNO_3$ solution.

## References

(1) Jagadeeswaraiah, K.; Balaraju, M.; Prasad, P. S.; Lingaiah, N. Selective Esterification of Glycerol to Bioadditives over Heteropoly Tungstate Supported on Cs-Containing Zirconia Catalysts. *Appl. Catal. Gen.* **2010**, *386* (1–2), 166–170.

(2) Bachrach, U.; Friedmann, A. Practical Procedures for the Purification of Bacterial Viruses. *Appl. Microbiol.* **1971**, *22* (4), 706–715.

(3) Bella, F.; Renzi, P.; Cavallo, C.; Gerbaldi, C. Caesium for Perovskite Solar Cells: An Overview. *Chem. Eur. J.* **2018**, *24* (47), 12183–12205.

(4) Kroeker, S.; Schuller, S.; Wren, J. E.; Greer, B. J.; Mesbah, A. 133 Cs and 23 Na MAS NMR Spectroscopy of Molybdate Crystallization in Model Nuclear Glasses. *J. Am. Ceram. Soc.* **2016**, *99* (5), 1557–1564.

(5) Gin, S.; Jollivet, P.; Tribet, M.; Peuget, S.; Schuller, S. Radionuclides Containment in Nuclear Glasses: An Overview. *Radiochim. Acta* **2017**, *105* (11), 927–959.

(6) Khalil, M.; Merz, E. Immobilization of Intermediate-Level Wastes in Geopolymers. *J. Nucl. Mater.* **1994**, *211* (2), 141–148.

(7) Hong, S.-Y.; Glasser, F. P. Alkali Binding in Cement Pastes. *Cem. Concr. Res.* **1999**, *29* (12), 1893–1903. https://doi.org/10.1016/S0008-8846(99)00187-8.

(8) Xu, H.; van Deventer, J. S. J. Microstructural Characterisation of Geopolymers Synthesized from Kaolinite/Stilbite Mixtures Using XRD, MAS-NMR, SEM/EDX, TEM/EDX and HREM. *Cem. Concr. Res.* **2002**, *32*, 1705–1716.

(9) Komnitsas, K.; Zaharaki, D. Geopolymerisation: A Review and Prospects for the Minerals Industry. *Miner. Eng.* **2007**, *20* (14), 1261–1277. https://doi.org/10.1016/j.mineng.2007.07.011.

(10) Ahn, M. K.; Iton, L. E. Solid-State Cesium-133 NMR Studies of Cations in Cesium/Lithium/Sodium Zeolite A: An Example of Cation Dynamics Involving Three Sites. *J. Phys. Chem.* **1991**, *95* (11), 4496–4500. https://doi.org/10.1021/j100164a059.

(11) Ashbrook, S. E.; Whittle, K. R.; Le Polles, L.; Farnan, I. Disorder and Dynamics in Pollucite from 133Cs and 27Al NMR. *J. Am. Ceram. Soc.* **2005**, *88* (6), 1575–1583. https://doi.org/10.1111/j.1551-2916.2005.00294.x.

(12) Ohkubo, T.; Okamoto, T.; Kawamura, K.; Guégan, R.; Deguchi, K.; Ohki, S.; Shimizu, T.; Tachi, Y.; Iwadate, Y. New Insights into the Cs Adsorption on Montmorillonite Clay from $^{133}$Cs Solid-State NMR and Density Functional Theory Calculations. *J. Phys. Chem. A* **2018**, *122* (48), 9326–9337. https://doi.org/10.1021/acs.jpca.8b07276.

(13) Arbel-Haddad, M.; Harnik, Y.; Schlosser, Y.; Goldbourt, A. Cesium Immobilization in Metakaolin-Based Geopolymers Elucidated by 133Cs Solid State NMR Spectroscopy. *J. Nucl. Mater.* **2022**, *562*, 153570. https://doi.org/10.1016/j.jnucmat.2022.153570.

(14) Lambregts, M. J.; Frank, S. M. Characterization of Cesium Containing Glass-Bonded Ceramic Waste Forms. *Microporous Mesoporous Mater.* **2003**, *64* (1–3), 1–9. https://doi.org/10.1016/S1387-1811(03)00486-4.

(15) Ibarra, I. A.; Lima, E.; Loera, S.; Bosch, P.; Bulbulian, S.; Lara, V. H. Cesium Leaching in CsA and CsX Zeolites: Use of Blocking Agents to Inhibit the Cesium Cation Mobility. *J. Phys. Chem. B* **2006**, *110* (42), 21086–21091. https://doi.org/10.1021/jp061926h.

(16) Norby, P.; Poshni, F. I.; Gualtieri, A. F.; Hanson, J. C.; Grey, C. P. Cation Migration in Zeolites: An in Situ Powder Diffraction and MAS NMR Study of the Structure of Zeolite Cs(Na)–Y during Dehydration. *J. Phys. Chem. B* **1998**, *102* (5), 839–856. https://doi.org/10.1021/jp9730398.

(17) Greer, B. J.; Kroeker, S. Characterisation of Heterogeneous Molybdate and Chromate Phase Assemblages in Model Nuclear Waste Glasses by Multinuclear Magnetic Resonance Spectroscopy. *Phys. Chem. Chem. Phys.* **2012**, *14* (20), 7375–7383.

(18) Pickard, C. J.; Mauri, F. All-Electron Magnetic Response with Pseudopotentials: NMR Chemical Shifts. *Phys. Rev. B* **2001**, *63* (24), 245101.

(19) Charpentier, T. The PAW/GIPAW Approach for Computing NMR Parameters: A New Dimension Added to NMR Study of Solids. *Solid State Nucl. Magn. Reson.* **2011**, *40* (1), 1–20.

(20) Bonhomme, C.; Gervais, C.; Babonneau, F.; Coelho, C.; Pourpoint, F.; Azais, T.; Ashbrook, S. E.; Griffin, J. M.; Yates, J. R.; Mauri, F. First-Principles Calculation of NMR Parameters Using the Gauge Including Projector Augmented Wave Method: A Chemist's Point of View. *Chem. Rev.* **2012**, *112* (11), 5733–5779.

(21) Schroeder, C.; Hansen, M. R.; Koller, H. Spatial Proximities between Brønsted Acid Sites, AlOH Groups, and Residual NH4+ Cations in Zeolites Mordenite and Ferrierite. *J. Phys. Chem. C* **2023**, *127* (1), 736–745.

(22) Dib, E.; Mintova, S.; Vayssilov, G. N.; Aleksandrov, H. A.; Carravetta, M. Chemical Shift Anisotropy: A Promising Parameter to Distinguish the 29Si NMR Peaks in Zeolites. *J. Phys. Chem. C* **2023**, *127* (22), 10792–10796.

(23) Hooper, R. W.; Ni, C.; Tkachuk, D. G.; He, Y.; Terskikh, V. V.; Veinot, J. G.; Michaelis, V. K. Exploring Structural Nuances in Germanium Halide Perovskites Using Solid-State 73Ge and 133Cs NMR Spectroscopy. *J. Phys. Chem. Lett.* **2022**, *13* (7), 1687–1696.

(24) Moon, C. J.; Park, J.; Im, H.; Ryu, H.; Choi, M. Y.; Kim, T. H.; Kim, J. Chemical Shift and Second-Order Quadrupolar Effects in the Solid-State 133Cs NMR Spectra of [Cs+ (Cryptand [2.2. 2])] X (X= I−, SCN−· H2O). *Bull. Korean Chem. Soc.* **2020**, *41* (7), 702–708.

(25) Toledano, P.; Knorr, K.; Ehm, L.; Depmeier, W. Phenomenological Theory of the Reconstructive Phase Transition between the NaCl and CsCl Structure Types. *Phys. Rev. B* **2003**, *67* (14), 144106.

(26) Posnjak, E.; Wyckoff, R. W. The Crystal Structures of the Alkali Halides. II. *J. Wash. Acad. Sci.* **1922**, *12* (10), 248–251.

(27) Mayer, J. E. Dispersion and Polarizability and the van Der Waals Potential in the Alkali Halides. *J. Chem. Phys.* **1933**, *1* (4), 270–279.

(28) London, F. The General Theory of Molecular Forces. *Trans. Faraday Soc.* **1937**, *33*, 8b–26.

(29) Zhang, F.; Gale, J.; Uberuaga, B.; Stanek, C.; Marks, N. Importance of Dispersion in Density Functional Calculations of Cesium Chloride and Its Related Halides. *Phys. Rev. B* **2013**, *88* (5), 054112.

(30) Perdew, J. P.; Burke, K.; Ernzerhof, M. Generalized Gradient Approximation Made Simple. *Phys. Rev. Lett.* **1996**, *77* (18), 3865.

(31) Köcher, S. S.; Schleker, P.; Graf, M.; Eichel, R.-A.; Reuter, K.; Granwehr, J.; Scheurer, C. Chemical Shift Reference Scale for Li Solid State NMR Derived by First-Principles DFT Calculations. *J. Magn. Reson.* **2018**, *297*, 33–41.

(32) Iwase, F. Ab Initio Calculations of Nuclear Quadrupole Resonance Frequencies in Trichloroacetyl Halides: A Comparison of DFT and Experimental Data. *Mater. Res. Express* **2020**, *7* (2), 025104.

(33) Reina, J. V.; Civaia, F.; Harper, A. F.; Scheurer, C.; Köcher, S. The EFG Rosetta Stone: Translating between DFT Calculations and Solid State NMR Experiments. *Faraday Discuss.* **2024**.

(34) Hartman, J. D.; Capistran, D. Predicting 51V Nuclear Magnetic Resonance Observables in Molecular Crystals. *Magn. Reson. Chem.* **2024**, *62* (6), 416–428.

(35) Park, J.-S. Comparison Study of Exchange-Correlation Functionals on Prediction of Ground States and Structural Properties. *Curr. Appl. Phys.* **2021**, *22*, 61–64.

(36) DiLabio, G. A.; Otero-de-la-Roza, A. Noncovalent Interactions in Density Functional Theory. *Rev. Comput. Chem.* **2016**, *29*, 1–97.

(37) Grønbech, T. B. E.; Tolborg, K.; Ceresoli, D.; Iversen, B. B. Anharmonic Motion and Aspherical Nuclear Probability Density Functions in Cesium Halides. *Phys. Rev. B* **2022**, *105* (10), 104113.

(38) Grimme, S. Accurate Description of van Der Waals Complexes by Density Functional Theory Including Empirical Corrections. *J. Comput. Chem.* **2004**, *25* (12), 1463–1473.

(39) Grimme, S. Semiempirical GGA-type Density Functional Constructed with a Long-range Dispersion Correction. *J. Comput. Chem.* **2006**, *27* (15), 1787–1799.

(40) Tkatchenko, A.; Scheffler, M. Accurate Molecular van Der Waals Interactions from Ground-State Electron Density and Free-Atom Reference Data. *Phys. Rev. Lett.* **2009**, *102* (7), 073005.

(41) Steinmann, S. N.; Csonka, G.; Corminboeuf, C. Unified Inter-and Intramolecular Dispersion Correction Formula for Generalized Gradient Approximation Density Functional Theory. *J. Chem. Theory Comput.* **2009**, *5* (11), 2950–2958.

(42) Steinmann, S. N.; Corminboeuf, C. Comprehensive Benchmarking of a Density-Dependent Dispersion Correction. *J. Chem. Theory Comput.* **2011**, *7* (11), 3567–3577.

(43) Sun, Y.; Zhang, Z.; Grigoryants, V. M.; Myers, W. K.; Liu, F.; Earle, K. A.; Freed, J. H.; Scholes, C. P. The Internal Dynamics of Mini c TAR DNA Probed by Electron Paramagnetic Resonance of Nitroxide Spin-Labels at the Lower Stem, the Loop, and the Bulge. *Biochemistry* **2012**, *51* (43), 8530–8541.

(44) Johnson, E. R. The Exchange-Hole Dipole Moment Dispersion Model. In *Non-covalent Interactions in Quantum Chemistry and Physics Theory and Applications*; de la Roza, A. O., DiLabio, G. A., Eds.; Elsevier, 2017; pp 169–194.

(45) Caldeweyher, E.; Ehlert, S.; Hansen, A.; Neugebauer, H.; Spicher, S.; Bannwarth, C.; Grimme, S. A Generally Applicable Atomic-Charge Dependent London Dispersion Correction. *J. Chem. Phys.* **2019**, *150* (15), 154122.

(46) Grimme, S.; Antony, J.; Ehrlich, S.; Krieg, H. A Consistent and Accurate Ab Initio Parametrization of Density Functional Dispersion Correction (DFT-D) for the 94 Elements H-Pu. *J. Chem. Phys.* **2010**, *132* (15), 154104.

(47) Ambrosetti, A.; Reilly, A. M.; DiStasio, R. A.; Tkatchenko, A. Long-Range Correlation Energy Calculated from Coupled Atomic Response Functions. *J. Chem. Phys.* **2014**, *140* (18).

(48) Becke, A. D.; Johnson, E. R. Exchange-Hole Dipole Moment and the Dispersion Interaction. *J. Chem. Phys.* **2005**, *122* (15), 154104.

(49) Becke, A. D.; Johnson, E. R. A Density-Functional Model of the Dispersion Interaction. *J. Chem. Phys.* **2005**, *123* (15).

(50) Otero-De-La-Roza, A.; Johnson, E. R. Van Der Waals Interactions in Solids Using the Exchange-Hole Dipole Moment Model. *J. Chem. Phys.* **2012**, *136* (17).

(51) Andersson, Y.; Langreth, D. C.; Lundqvist, B. I. Van Der Waals Interactions in Density-Functional Theory. *Phys. Rev. Lett.* **1996**, *76* (1), 102–105. https://doi.org/10.1103/PhysRevLett.76.102.

(52) Dobson, J. F.; Dinte, B. P. Constraint Satisfaction in Local and Gradient Susceptibility Approximations: Application to a van Der Waals Density Functional. *Phys. Rev. Lett.* **1996**, *76* (11), 1780.

(53) Dion, M.; Rydberg, H.; Schröder, E.; Langreth, D. C.; Lundqvist, B. I. Van Der Waals Density Functional for General Geometries. *Phys. Rev. Lett.* **2004**, *92* (24), 246401.

(54) Vydrov, O. A.; Van Voorhis, T. Nonlocal van Der Waals Density Functional Made Simple. *Phys. Rev. Lett.* **2009**, *103* (6), 063004.

(55) Langreth, D.; Lundqvist, B. I.; Chakarova-Käck, S. D.; Cooper, V.; Dion, M.; Hyldgaard, P.; Kelkkanen, A.; Kleis, J.; Kong, L.; Li, S. A Density Functional for Sparse Matter. *J. Phys. Condens. Matter* **2009**, *21* (8), 084203.

(56) Lee, K.; Murray, É. D.; Kong, L.; Lundqvist, B. I.; Langreth, D. C. Higher-Accuracy van Der Waals Density Functional. *Phys. Rev. B* **2010**, *82* (8), 081101.

(57) Vydrov, O. A.; Van Voorhis, T. Nonlocal van Der Waals Density Functional: The Simpler the Better. *J. Chem. Phys.* **2010**, *133* (24), 244103.

(58) Sabatini, R.; Gorni, T.; De Gironcoli, S. Nonlocal van Der Waals Density Functional Made Simple and Efficient. *Phys. Rev. B* **2013**, *87* (4), 041108.

(59) Chakraborty, D.; Berland, K.; Thonhauser, T. Next-Generation Nonlocal van Der Waals Density Functional. *J. Chem. Theory Comput.* **2020**, *16* (9), 5893–5911.

(60) Lin, I.-C.; Coutinho-Neto, M. D.; Felsenheimer, C.; von Lilienfeld, O. A.; Tavernelli, I.; Rothlisberger, U. Library of Dispersion-Corrected Atom-Centered Potentials for Generalized Gradient Approximation Functionals: Elements H, C, N, O, He, Ne, Ar, and Kr. *Phys. Rev. B* **2007**, *75* (20), 205131.

(61) Mackie, I. D.; DiLabio, G. A. Interactions in Large, Polyaromatic Hydrocarbon Dimers: Application of Density Functional Theory with Dispersion Corrections. *J. Phys. Chem. A* **2008**, *112* (43), 10968–10976.

(62) DiLabio, G. A. Accurate Treatment of van Der Waals Interactions Using Standard Density Functional Theory Methods with Effective Core-Type Potentials: Application to Carbon-Containing Dimers. *Chem. Phys. Lett.* **2008**, *455* (4–6), 348–353.

(63) Torres, E.; DiLabio, G. A. A (Nearly) Universally Applicable Method for Modeling Noncovalent Interactions Using B3LYP. *J. Phys. Chem. Lett.* **2012**, *3* (13), 1738–1744.

(64) Zhao, Y.; Schultz, N. E.; Truhlar, D. G. Exchange-Correlation Functional with Broad Accuracy for Metallic and Nonmetallic Compounds, Kinetics, and Noncovalent Interactions. *J. Chem. Phys.* **2005**, *123* (16), 161103.

(65) Zhao, Y.; Truhlar, D. G. The M06 Suite of Density Functionals for Main Group Thermochemistry, Thermochemical Kinetics, Noncovalent Interactions, Excited States, and Transition Elements: Two New Functionals and Systematic Testing of Four M06-Class Functionals and 12 Other Functionals. *Theor. Chem. Acc.* **2008**, *120* (1), 215–241.

(66) Peverati, R.; Truhlar, D. G. Exchange–Correlation Functional with Good Accuracy for Both Structural and Energetic Properties While Depending Only on the Density and Its Gradient. *J. Chem. Theory Comput.* **2012**, *8* (7), 2310–2319.

(67) Perdew, J. P.; Kurth, S.; Zupan, A.; Blaha, P. Accurate Density Functional with Correct Formal Properties: A Step beyond the Generalized Gradient Approximation. *Phys. Rev. Lett.* **1999**, *82* (12), 2544.

(68) Sun, J.; Ruzsinszky, A.; Perdew, J. P. Strongly Constrained and Appropriately Normed Semilocal Density Functional. *Phys. Rev. Lett.* **2015**, *115* (3), 036402.

(69) Ohkubo, T.; Takei, A.; Tachi, Y.; Fukatsu, Y.; Deguchi, K.; Ohki, S.; Shimizu, T. New Approach To Understanding the Experimental 133Cs NMR Chemical Shift of Clay Minerals via Machine Learning and DFT-GIPAW Calculations. *J. Phys. Chem. A* **2023**, *127* (4), 973–986.

(70) Giannozzi, P.; Baroni, S.; Bonini, N.; Calandra, M.; Car, R.; Cavazzoni, C.; Ceresoli, D.; Chiarotti, G. L.; Cococcioni, M.; Dabo, I. QUANTUM ESPRESSO: A Modular and Open-Source Software Project for Quantum Simulations of Materials. *J. Phys. Condens. Matter* **2009**, *21* (39), 395502.

(71) Giannozzi, P.; Andreussi, O.; Brumme, T.; Bunau, O.; Nardelli, M. B.; Calandra, M.; Car, R.; Cavazzoni, C.; Ceresoli, D.; Cococcioni, M. Advanced Capabilities for Materials Modelling with Quantum ESPRESSO. *J. Phys. Condens. Matter* **2017**, *29* (46), 465901.

(72) Dal Corso, A. Pseudopotentials Periodic Table: From H to Pu. *Comput. Mater. Sci.* **2014**, *95*, 337–350.

(73) Hammer, B.; Hansen, L. B.; Nørskov, J. K. Improved Adsorption Energetics within Density-Functional Theory Using Revised Perdew-Burke-Ernzerhof Functionals. *Phys. Rev. B* **1999**, *59* (11), 7413.

(74) Zhang, Y.; Yang, W. Comment on "Generalized Gradient Approximation Made Simple." *Phys. Rev. Lett.* **1998**, *80* (4), 890.

(75) Holmes, S. T.; Alkan, F.; Iuliucci, R. J.; Mueller, K. T.; Dybowski, C. Analysis of the Bond-valence Method for Calculating 29Si and 31P Magnetic Shielding in Covalent Network Solids. *J. Comput. Chem.* **2016**, *37* (18), 1704–1710.

(76) Holmes, S. T.; Bai, S.; Iuliucci, R. J.; Mueller, K. T.; Dybowski, C. Calculations of Solid-state 43Ca NMR Parameters: A Comparison of Periodic and Cluster Approaches and an Evaluation of DFT Functionals. *J. Comput. Chem.* **2017**, *38* (13), 949–956.

(77) Perdew, J. P.; Ruzsinszky, A.; Csonka, G. I.; Vydrov, O. A.; Scuseria, G. E.; Constantin, L. A.; Zhou, X.; Burke, K. Restoring the Density-Gradient Expansion for Exchange in Solids and Surfaces. *Phys. Rev. Lett.* **2008**, *100* (13), 136406.

(78) Becke, A. D. On the Large-gradient Behavior of the Density Functional Exchange Energy. *J. Chem. Phys.* **1986**, *85* (12), 7184–7187. https://doi.org/10.1063/1.451353.

(79) Becke, A. D.; Johnson, E. R. Exchange-Hole Dipole Moment and the Dispersion Interaction Revisited. *J. Chem. Phys.* **2007**, *127* (15).

(80) Tao, J.; Zheng, F.; Gebhardt, J.; Perdew, J. P.; Rappe, A. M. Screened van Der Waals Correction to Density Functional Theory for Solids. *Phys. Rev. Mater.* **2017**, *1* (2), 020802.

(81) Berland, K.; Chakraborty, D.; Thonhauser, T. Van Der Waals Density Functional with Corrected C 6 Coefficients. *Phys. Rev. B* **2019**, *99* (19), 195418.

(82) Hamada, I. Van Der Waals Density Functional Made Accurate. *Phys. Rev. B* **2014**, *89* (12), 121103.

(83) Holmes, S. T.; Schurko, R. W. Refining Crystal Structures with Quadrupolar NMR and Dispersion-Corrected Density Functional Theory. *J. Phys. Chem. C* **2018**, *122* (3), 1809–1820.

(84) Yang, J. H.; Kitchaev, D. A.; Ceder, G. Rationalizing Accurate Structure Prediction in the Meta-GGA SCAN Functional. *Phys. Rev. B* **2019**, *100* (3), 035132.

(85) Anatole von Lilienfeld, O.; Tkatchenko, A. Two-and Three-Body Interatomic Dispersion Energy Contributions to Binding in Molecules and Solids. *J. Chem. Phys.* **2010**, *132* (23).

(86) Pyper, N. The Cohesive Energetics of Solid Cesium Chloride. *J. Chem. Phys.* **2003**, *118* (5), 2308–2324.

(87) Aguado, A. Comment on "The Cohesive Energetics of Solid Cesium Chloride"[J. Chem. Phys. 118, 2308 (2003)]. *J. Chem. Phys.* **2003**, *119* (16), 8765–8766.

(88) Otero-de-la-Roza, A.; Johnson, E. R. Application of XDM to Ionic Solids: The Importance of Dispersion for Bulk Moduli and Crystal Geometries. *J. Chem. Phys.* **2020**, *153* (5), 054121.

(89) Czernek, J.; Brus, J. Describing the Anisotropic 133Cs Solid State NMR Interactions in Cesium Chromate. *Chem. Phys. Lett.* **2017**, *684*, 8–13.

(90) Halliday, J.; Hill, H.; Richards, R. Solvent Isotope Shifts of Caesium Resonances in Dilute Salt Solutions. *J. Chem. Soc. Chem. Commun.* **1969**, No. 5, 219–220.

(91) Haase, A.; Kerber, M.; Kessler, D.; Kronenbitter, J.; Krüger, H.; Lutz, O.; Müller, M.; Nolle, A. Nuclear Magnetic Shielding and Quadrupole Coupling of 133Cs in Cesium Salt Powders. *Z. Für Naturforschung A* **1977**, *32* (9), 952–956.

(92) Mooibroek, S.; Wasylishen, R. E.; Dickson, R.; Facey, G.; Pettitt, B. A. Simultaneous Observation of Shielding Anisotropies and Quadrupolar Splittings in Solid State 133Cs NMR Spectra. *J. Magn. Reson. 1969* **1986**, *66* (3), 542–545.

(93) Kolganov, A. A.; Gabrienko, A. A.; Chernyshov, I. Y.; Stepanov, A. G.; Pidko, E. A. The Accuracy Challenge of the DFT-Based Molecular Assignment of 13 C MAS NMR Characterization of Surface Intermediates in Zeolite Catalysis. *Phys. Chem. Chem. Phys.* **2020**, *22* (41), 24004–24013.

(94) Dietrich, B.; Lehn, J.; Sauvage, J. Les Cryptates. *Tetrahedron Lett.* **1969**, *10* (34), 2889–2892.

(95) Kubicki, D. J.; Prochowicz, D.; Hofstetter, A.; Zakeeruddin, S. M.; Grätzel, M.; Emsley, L. Phase Segregation in Cs-, Rb-and K-Doped Mixed-Cation (MA) x (FA) 1–x PbI3 Hybrid Perovskites from Solid-State NMR. *J. Am. Chem. Soc.* **2017**, *139* (40), 14173–14180.

(96) Thiele, G.; Messer, D. S-Thiocyanato-und N-Isothiocyanato-Bindungsisomerie in Den Kristallstrukturen von RbCd (SCN) 3 Und CsCd (SCN) 3. *Z. Für Anorg. Allg. Chem.* **1980**, *464* (1), 255–267.

(97) Behrens, U.; Hoffmann, F.; Olbrich, F. Solid-State Structures of Base-Free Lithium and Sodium Tetraphenylborates at Room and Low Temperature: Comparison with the Higher Homologues MB (C6H5) 4 (M= K, Rb, Cs). *Organometallics* **2012**, *31* (3), 905–913.

(98) Krogh-Moe, J. The Crystal Structure of Cesium Triborate, Cs2O. 3B2O3. *Acta Crystallogr.* **1960**, *13* (11), 889–892.

(99) Rakhmatullin, A.; Allix, M.; Polovov, I. B.; Maltsev, D.; Chukin, A. V.; Bakirov, R.; Bessada, C. Combining Solid State NMR, Powder X-Ray Diffraction, and DFT Calculations for CsSc3F10 Structure Determination. *J. Alloys Compd.* **2019**, *787*, 1349–1355.

(100) Trots, D.; Myagkota, S. High-Temperature Structural Evolution of Caesium and Rubidium Triiodoplumbates. *J. Phys. Chem. Solids* **2008**, *69* (10), 2520–2526.

(101) Fadla, M. A.; Bentria, B.; Dahame, T.; Benghia, A. First-Principles Investigation on the Stability and Material Properties of All-Inorganic Cesium Lead Iodide Perovskites CsPbI3 Polymorphs. *Phys. B Condens. Matter* **2020**, *585*, 412118.

(102) Tran, F.; Kalantari, L.; Traoré, B.; Rocquefelte, X.; Blaha, P. Nonlocal van Der Waals Functionals for Solids: Choosing an Appropriate One. *Phys. Rev. Mater.* **2019**, *3* (6), 063602.

(103) Del Grande, R. R.; Menezes, M. G.; Capaz, R. B. Layer Breathing and Shear Modes in Multilayer Graphene: A DFT-vdW Study. *J. Phys. Condens. Matter* **2019**, *31* (29), 295301.

(104) Takano, Y.; Kobayashi, N.; Morikawa, Y. Computational Study on Atomic Structures, Electronic Properties, and Chemical Reactions at Surfaces and Interfaces and in Biomaterials. *J. Phys. Soc. Jpn.* **2018**, *87* (6), 061013.

(105) Abidin, A. F. Z.; Hamada, I. Interaction of Water with Nitrogen-Doped Graphene. *Phys. Rev. B* **2022**, *105* (7), 075416.

(106) Bernasconi, D.; Bordignon, S.; Rossi, F.; Priola, E.; Nervi, C.; Gobetto, R.; Voinovich, D.; Hasa, D.; Duong, N. T.; Nishiyama, Y. Selective Synthesis of a Salt and a Cocrystal of the Ethionamide–Salicylic Acid System. *Cryst. Growth Des.* **2019**, *20* (2), 906–915.

# Supplementary Information:

# Prediction of NMR Parameters and geometry in $^{133}$Cs-containing compounds using Density Functional Theory

N. Manukovsky[1], N. Vaisleib[1], M. Arbel-Haddad[2], A. Goldbourt[1]

[1]School of Chemistry, Tel Aviv University, Ramat Aviv 6997801, Tel Aviv, Israel
[2]Nuclear Research Center Negev, PO Box 9001, Beer Sheva 84901, Israel

## Computational parameters

Table S1 lists the energy cutoff values and Monkhorst-Pack grid dimensions used in each calculation.

*Table S1: energy cutoff and grid sizes used for the calculations.*

| system | Ecut [Ry] | kgrid xyz |
|---|---|---|
| CsF | 100 | 12 12 12 |
| CsCl | 80 | 8 8 8 |
| CsBr | 100 | 8 8 8 |
| CsI | 100 | 8 8 8 |
| $Cs_2CrO_4$ | 80 | 5 6 4 |
| $CsClO_4$ | 100 | 7 6 4 |
| $Cs_2SO_4$ | 100 | 8 6 5 |
| $CsVO_3$ | 110 | 10 7 10 |
| $CsGeCl_3$ | 100 | 12 12 12 |
| $CsGeBr_3$ | 100 | 12 12 12 |
| $CsGeI_3$ | 100 | 12 12 12 |
| $CsCd(SCN)_3$ | 70 | 4 6 3 |
| $CsBPh_4$ | 70 | 4 4 4 |
| $[Cs^+(C222)]I^-$ | 90 | 3 3 3 |
| $CsB_3O_5$ | 90 | 5 4 3 |
| $CsSc_3F_{10}$ | 90 | 5 5 6 |
| $CsPbI_3$, hexagonal | 110 | 4 8 2 |
| $CsPbI_3$, tetragonal | 110 | 5 5 3 |

## Crystal structures and unit cell volumes

Table S2 specifies the experimental and computational unit cell volumes, as well as the references for the crystal structures used as input for the structural optimizations.

*Table S2: Experimental and calculated unit cell volumes [Å³] of the tested materials*

| | Exp | ref | PBE | rVV10 | vdW-DF3-opt1 | vdW-DF-C6 | rev-vdW-DF2 | PBE+XDM | PBE+D3 | PBE +D2 Zhang | rPBE+D2 Zhang | rPBE+D2* | rPBE+D2* Zhang | revPBE+D3 | PBEsol+D2 Zhang | PBEsol+D3 | B86bPBE+XDM |
|---|---|---|---|---|---|---|---|---|---|---|---|---|---|---|---|---|---|
| CsF | 54.22 | 1 | 57.11 | 53.52 | 52.02 | 53.56 | 54.12 | 41.56 | 56.28 | 51.87 | 56.55 | 36.03 | 55.24 | 61.19 | 48.10 | 51.99 | 47.30 |
| CsCl | 70.09 | 2 | 74.24 | 68.54 | 66.25 | 68.22 | 68.92 | 50.97 | 74.36 | 67.51 | 72.80 | 48.64 | 72.80 | 79.42 | 63.21 | 67.32 | 58.74 |
| CsBr | 78.73 | 2 | 83.87 | 78.02 | 78.24 | 78.75 | 78.68 | 57.86 | 84.48 | 75.92 | 78.73 | 55.50 | 81.50 | 90.59 | 70.99 | 75.97 | 66.37 |
| CsI | 95.24 | 2 | 101.49 | 93.22 | 90.17 | 92.81 | 93.87 | 68.7 2 | 102.04 | 89.09 | 95.27 | 68.18 | 95.27 | 111.48 | 83.86 | 90.65 | 79.61 |
| $Cs_2CrO_4$ | 594.61 | 3 | 630.09 | 587.09 | 572.47 | 585.99 | 591.49 | 492.05 | 620.04 | 574.66 | 618.24 | 463.99 | 616.76 | 658.95 | 536.45 | 573.54 | 546.63 |
| $CsClO_4$ | 458.28 | 4 | 499.01 | 447.12 | 438.86 | 449.31 | 455.34 | 401.77 | 480.19 | 455.49 | 497.34 | 417.34 | 497.02 | 499.35 | 425.70 | 450.57 | 429.84 |
| $Cs_2SO_4$ | 563.72 | 5 | 602.80 | 559.22 | 544.48 | 556.51 | 561.45 | 475.88 | 592.37 | 555.57 | 594.91 | 444.64 | 586.52 | 625.40 | 522.00 | 549.93 | 475.88 |
| $CsVO_3$ | 382.21 | 6 | 418.31 | 381.22 | 371.36 | 380.15 | 384.94 | 356.19 | 408.41 | 374.63 | 409.59 | 335.44 | 408.01 | 427.01 | 349.24 | 375.67 | 354.54 |
| $CsGeCl_3$ | 161.43 | 7 | 169.12 | 152.18 | 145.96 | 150.68 | 152.20 | 133.52 | 163.34 | 147.76 | 171.30 | 145.74 | 171.76 | 167.85 | 134.47 | 146.70 | 140.68 |
| $CsGeBr_3$ | 179.98 | 7 | 189.43 | 174.38 | 165.74 | 170.85 | 173.06 | 153.34 | 181.75 | 165.04 | 185.94 | 160.81 | 186.76 | 185.40 | 153.50 | 164.10 | 159.23 |
| $CsGeI_3$ | 215.37 | 7 | 228.01 | 214.47 | 203.09 | 208.46 | 211.53 | 184.93 | 219.52 | 199.25 | 218.47 | 193.80 | 216.10 | 227.74 | 185.48 | 200.71 | 194.96 |
| $CsCd(SCN)_3$ | 967.6 | 8 | 1077.6 | 946.4 | | 938.9 | 951.3 | | 1010.1 | | | | | 1038.0 | | 927.0 | |
| $CsBPh_4$ | 536.4 | 9 | 609.7 | 502.2 | | 500.9 | 510.7 | | 545.1 | | | | | 602.7 | | 503.0 | |
| [Cs⁺(C222)]I⁻ | 1235.4 | 10 | 1438.2 | 1174.8 | | 1183.6 | 1208.8 | | 1234.5 | | | | | 1215.7 | | 1144.4 | |
| $CsB_3O_5$ | 480.6 | 11 | 522.7 | 488.6 | | 484.7 | 489.2 | | 505.4 | | | | | 516.9 | | 480.0 | |
| $CsSc_3F_{10}$ | 417.2 | 12 | 433.8 | 416.9 | | 421.5 | 423.2 | | 430.6 | | | | | 440.7 | | 413.6 | |
| $CsPbI_3$, hexagonal | 892.7 | 14 | 964.0 | 895.0 | | 883.9 | 897.5 | | 934.3 | | | | | 975.9 | | 855.1 | |
| $CsPbI_3$, tetragonal | 947.3 | 15 | 1005.3 | 939.5 | | 929.9 | 947.5 | | 983.8 | | | | | 1011.9 | | 913.5 | |

## Additional error metrics

Figure S1 shows additional statistical parameters of the DFT calculations.

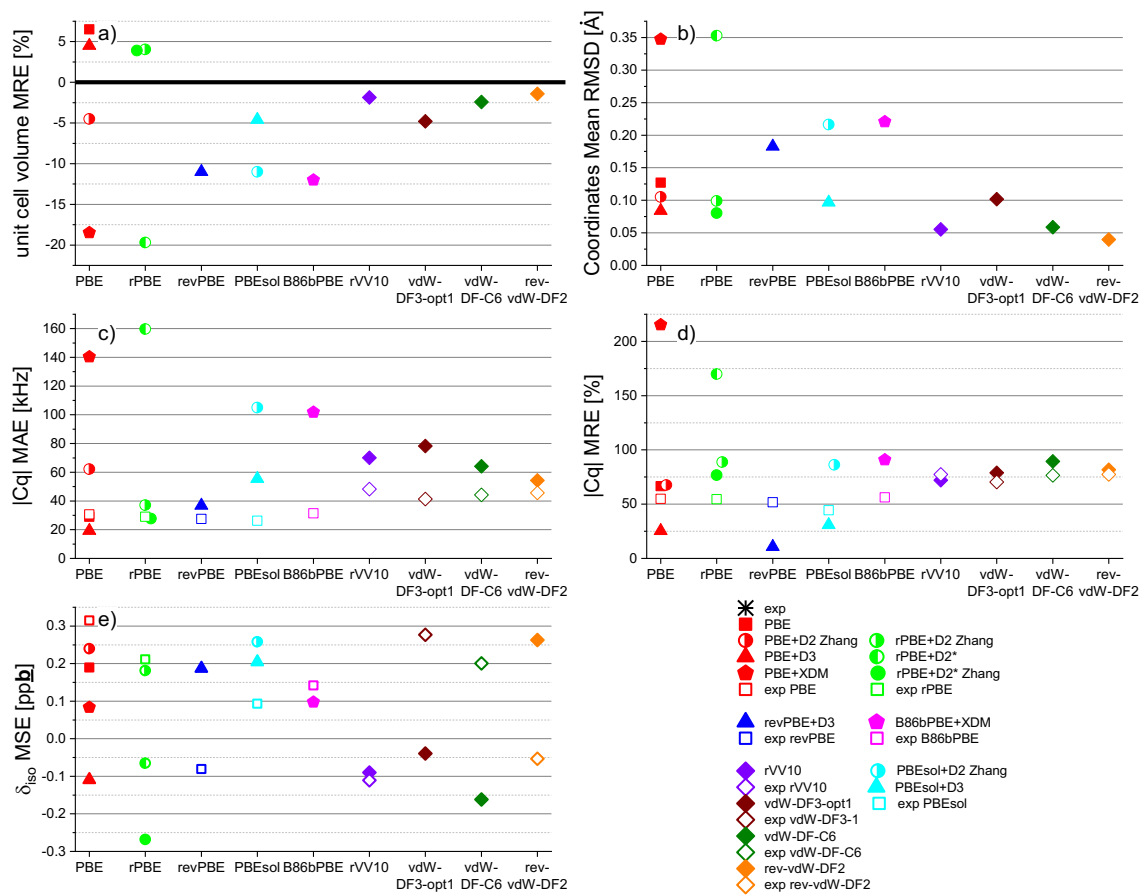

Figure S1: Error metrics of DFT functionals: a) Mean Relative Error of the unit cell volume; b) Mean RMSD of the atom coordinates; c) Mean Absolute Error and d) Mean Relative Error of the quadrupolar coupling constant; e) Mean Signed Error of the isotropic chemical shift, in ppb.

## Quadrupolar coupling constants

Table S3 – Table S6 specify the experimental and computational quadrupolar coupling parameters of the tested materials.

*Table S3: Experimental quadrupolar coupling constants of the tested materials*

| | Experimental | | |
|---|---|---|---|
| | $|C_q|$ [kHz] | η | ref |
| **CsF** | 0 | 0 | |
| **CsCl** | 0 | 0 | |
| **CsBr** | 0 | 0 | |
| **CsI** | 0 | 0 | |
| **$Cs_2CrO_4$ I** | 365 | 0.56 | [16] |
| | 376 | 0.52 | [17] |
| | 373 | 0.55 | [18] |
| **$Cs_2CrO_4$ II** | 142 | 0.11 | [16] |
| | 138 | 0.15 | [17] |
| | 142 | 0.15 | [18] |
| **$CsClO_4$** | 133.6 | 0.11 | [16] |
| **$Cs_2SO_4$ I** | 20 | 0.27 | [16] |
| **$Cs_2SO_4$ II** | 261 | 0.01 | [16] |
| **$CsVO_3$** | 225 | 0.47 | [16] |
| **$CsGeCl_3$** | 12 | <0.10 | [7] |
| **$CsGeBr_3$** | 52 | 0 | [7] |
| | 50 | 0 | |
| **$CsGeI_3$** | 55 | 0.05 | [7] |
| | 52 | 0.05 | |
| **$CsCd(SCN)_3$** | 148 | 0.98 | [19] |
| **$CsBPh_4$** | 335 | NA | [20] |
| **[$Cs^+$(Cryptand[2.2.2])]I$^-$** | 1047 | 0.73 | [10] |
| **$CsB_3O_5$** | 380 | NA | [21] |
| **$CsSc_3F_{10}$** | NA | NA | |
| **$CsPbI_3$, hexagonal** | NA | NA | |
| **$CsPbI_3$, tetragonal** | NA | NA | |

*Table S4: Calculated quadrupolar coupling constants – PBE and rPBE functionals*

|  | PBE | | Exp PBE | | PBE + D2 Zhang | | PBE+D3 | | PBE+XDM | | rPBE + D2 Zhang | | rPBE +D2* | | rPBE +D2* Zhang | | exp rPBE | |
|---|---|---|---|---|---|---|---|---|---|---|---|---|---|---|---|---|---|---|
|  | Cq [kHz] | η | Cq [kHz] | η | Cq [kHz] | η | Cq [kHz] | η | Cq [kHz] | η | Cq [kHz] | η | Cq [kHz] | η | Cq [kHz] | η | Cq [kHz] | η |
| Cs$_2$CrO$_4$ I | 391 | 0.6 | 456 | 0.6 | 487 | 0.4 | 396 | 0.6 | 725 | 0.4 | 373 | 0.6 | 544 | 0.8 | 393 | 0.6 | 445 | 0.5 |
| Cs$_2$CrO$_4$ II | -136 | 0.5 | -152 | 0.2 | -183 | 0.4 | -142 | 0.0 | -193 | 0.1 | -154 | 0.3 | -450 | 0.0 | -141 | 0.3 | -153 | 0.3 |
| CsClO$_4$ | 156 | 0.3 | 194 | 0.8 | 304 | 0.1 | 203 | 0.2 | 430 | 0.2 | 218 | 0.3 | 406 | 0.1 | 155 | 0.3 | 189 | 0.8 |
| Cs$_2$SO$_4$ I | 278 | 0.3 | 289 | 0.0 | 293 | 0.2 | 269 | 0.1 | 359 | 0.2 | 275 | 0.2 | 342 | 0.2 | 274 | 0.3 | 281 | 0.0 |
| Cs$_2$SO$_4$ II | 82 | 0.3 | 64 | 0.6 | 116 | 0.5 | 43 | 0.1 | 242 | 0.2 | 79 | 0.5 | 212 | 0.6 | 84 | 0.3 | 62 | 0.7 |
| CsVO$_3$ | 140 | 0.2 | 238 | 0.3 | 266 | 0.3 | 195 | 0.8 | 373 | 0.8 | 146 | 0.5 | 571 | 1.0 | 159 | 0.5 | 244 | 0.1 |
| CsGeCl$_3$ | 41 | 0.0 | 28 | 0.0 | 11 | 0.0 | 20 | 0.0 | 53 | 0.0 | 56 | 0.0 | 8 | 0.0 | 48 | 0.0 | 29 | 0.0 |
| CsGeBr$_3$ | 60 | 0.0 | 56 | 0.0 | 20 | 0.0 | 43 | 0.0 | 17 | 0.0 | 73 | 0.0 | 17 | 0.0 | 64 | 0.0 | 57 | 0.0 |
| CsGeI$_3$ | 66 | 0.0 | 66 | 0.0 | 25 | 0.0 | 54 | 0.0 | 70 | 0.0 | 69 | 0.0 | 29 | 0.0 | 67 | 0.0 | 68 | 0.0 |
| CsCd(SCN)$_3$ | 217 | 0.6 |  |  |  |  | 160 | 1.0 |  |  |  |  |  |  |  |  |  |  |
| CsBPh$_4$ | 258 | 0.0 |  |  |  |  | 275 | 0.0 |  |  |  |  |  |  |  |  |  |  |
| [Cs$^+$(C222)]I$^-$ | 959 | 0.7 |  |  |  |  | 982 | 0.9 |  |  |  |  |  |  |  |  |  |  |
| CsB$_3$O$_5$ | -465 | 0.7 |  |  |  |  | -436 | 0.5 |  |  |  |  |  |  |  |  |  |  |
| CsSc$_3$F$_{10}$ | -431 | 0.8 |  |  |  |  | -432 | 0.8 |  |  |  |  |  |  |  |  |  |  |

*Table S5: Calculated quadrupolar coupling constants – additional semilocal functionals*

|  | revPBE + D3 | | exp revPBE | | PBEsol + D2 Zhang | | PBEsol + D3 | | exp PBEsol | | B86bPBE + XDM | | exp B86bPBE | |
|---|---|---|---|---|---|---|---|---|---|---|---|---|---|---|
|  | Cq [kHz] | η | Cq [kHz] | η | Cq [kHz] | η | Cq [kHz] | η | Cq [kHz] | η | Cq [kHz] | η | Cq [kHz] | η |
| Cs$_2$CrO$_4$ I | 276 | 0.8 | 444 | 0.5 | 616 | 0.3 | 506 | 0.6 | 451 | 0.6 | 573 | 0.6 | 458 | 0.6 |
| Cs$_2$CrO$_4$ II | -136 | 0.6 | -153 | 0.3 | -184 | 0.4 | -155 | 0.0 | -150 | 0.1 | -169 | 0.1 | -153 | 0.2 |
| CsClO$_4$ | 223 | 0.1 | 189 | 0.8 | 366 | 0.0 | 242 | 0.3 | 189 | 0.7 | 326 | 0.3 | 195 | 0.8 |
| Cs$_2$SO$_4$ I | 258 | 0.0 | 282 | 0.0 | 327 | 0.1 | 290 | 0.0 | 291 | 0.0 | 323 | 0.2 | 290 | 0.0 |
| Cs$_2$SO$_4$ II | 30 | 0.3 | 62 | 0.7 | 148 | 0.3 | 75 | 0.5 | 58 | 0.7 | 157 | 0.2 | 66 | 0.6 |
| CsVO$_3$ | 116 | 0.8 | 237 | 0.2 | 350 | 0.2 | 310 | 0.8 | 221 | 0.5 | 425 | 0.4 | 239 | 0.3 |
| CsGeCl$_3$ | 20 | 0.0 | 27 | 0.0 | 3 | 0.0 | 3 | 0.0 | 25 | 0.0 | 3 | 0.0 | 28 | 0.0 |
| CsGeBr$_3$ | 41 | 0.0 | 55 | 0.0 | 3 | 0.0 | 15 | 0.0 | 50 | 0.0 | 0 | 0.0 | 56 | 0.0 |
| CsGeI$_3$ | 57 | 0.0 | 65 | 0.0 | 6 | 0.0 | 28 | 0.0 | 59 | 0.0 | 18 | 0.0 | 66 | 0.0 |
| CsCd(SCN)$_3$ | 133 | 0.9 |  |  |  |  | 210 | 0.6 |  |  |  |  |  |  |
| CsBPh$_4$ | 215 | 0.0 |  |  |  |  | 296 | 0.0 |  |  |  |  |  |  |
| [Cs$^+$(Cryptand[2.2.2])]I$^-$ | 952 | 0.9 |  |  |  |  | 1020 | 0.9 |  |  |  |  |  |  |
| CsB$_3$O$_5$ | -335 | 0.7 |  |  |  |  | -499 | 0.7 |  |  |  |  |  |  |
| CsSc$_3$F$_{10}$ | -386 | 0.9 |  |  |  |  | -469 | 0.7 |  |  |  |  |  |  |

*Table S6: Calculated quadrupolar coupling constants – nonlocal functionals*

|  | rVV10 |  | exp rVV10 |  | vdW-DF3-opt1 |  | exp vdW-DF3-opt1 |  | vdW-DF-C6 |  | exp vdW-DF-C6 |  | rev-vdW-DF2 |  | exp rev-vdW-DF2 |  |
| --- | --- | --- | --- | --- | --- | --- | --- | --- | --- | --- | --- | --- | --- | --- | --- | --- |
|  | Cq [kHz] | η | Cq [kHz] | η | Cq [kHz] | η | Cq [kHz] | η | Cq [kHz] | η | Cq [kHz] | η | Cq [kHz] | η | Cq [kHz] | η |
| $Cs_2CrO_4$ I | 529 | 0.6 | 500 | 0.6 | 539 | 0.6 | 479 | 0.6 | 511 | 0.6 | 488 | 0.6 | 496 | 0.6 | 486 | 0.6 |
| $Cs_2CrO_4$ II | -158 | 0.2 | -154 | 0.3 | -158 | 0.3 | -153 | 0.3 | -155 | 0.3 | -154 | 0.3 | -152 | 0.3 | -154 | 0.3 |
| $CsClO_4$ | 300 | 0.2 | 218 | 0.7 | 310 | 0.3 | 210 | 0.8 | 289 | 0.2 | 194 | 0.8 | 259 | 0.3 | 214 | 0.8 |
| $Cs_2SO_4$ I | 319 | 0.2 | 309 | 0.1 | 322 | 0.2 | 301 | 0.1 | 315 | 0.2 | 303 | 0.1 | 310 | 0.2 | 302 | 0.1 |
| $Cs_2SO_4$ II | 104 | 0.4 | 82 | 0.5 | 120 | 0.5 | 76 | 0.6 | 105 | 0.5 | 80 | 0.6 | 99 | 0.5 | 80 | 0.6 |
| $CsVO_3$ | 354 | 0.5 | 272 | 0.2 | 358 | 0.5 | 257 | 0.3 | 323 | 0.5 | 270 | 0.2 | 296 | 0.4 | 268 | 0.2 |
| $CsGeCl_3$ | 8 | 0.0 | 31 | 0.0 | 12 | 0.0 | 31 | 0.0 | 28 | 0.0 | 33 | 0.0 | 28 | 0.0 | 33 | 0.0 |
| $CsGeBr_3$ | 37 | 0.0 | 63 | 0.0 | 27 | 0.0 | 61 | 0.0 | 41 | 0.0 | 65 | 0.0 | 44 | 0.0 | 65 | 0.0 |
| $CsGeI_3$ | 54 | 0.0 | 73 | 0.0 | 29 | 0.0 | 72 | 0.0 | 58 | 0.0 | 77 | 0.0 | 61 | 0.0 | 77 | 0.0 |
| $CsCd(SCN)_3$ | -244 | 0.6 |  |  |  |  |  |  | -255 | 0.6 |  |  | -245 | 0.7 |  |  |
| $CsBPh_4$ | 327 | 0.0 |  |  |  |  |  |  | 344 | 0.0 |  |  | 347 | 0.0 |  |  |
| [$Cs^+$(Cryptand[2.2.2])]$I^-$ | 1069 | 0.9 |  |  |  |  |  |  | 1072 | 1.0 |  |  | 1071 | 0.9 |  |  |
| $CsB_3O_5$ | -541 | 0.5 |  |  |  |  |  |  | -494 | 0.8 |  |  | -485 | 0.7 |  |  |
| $CsSc_3F_{10}$ | -555 | 0.9 |  |  |  |  |  |  | -519 | 0.6 |  |  | -515 | 0.6 |  |  |

## Scaling the quadrupolar coupling constants

Figure S2 and Table S7 show a fairly linear, though noisy, correlation between the computational |Cq| obtained by various functionals (shown are the best performing functionals in terms of Cq). Similar trends were reported for $^{133}$Cs[22] and for several other half-integer quadrupolar nuclei[23].

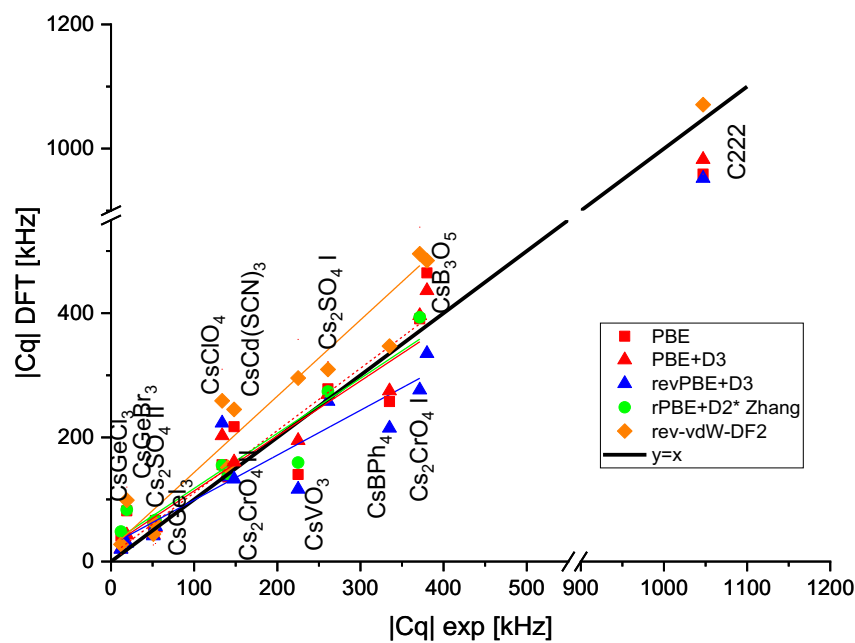

Figure S2: Calculated vs experimental quadrupolar coupling constants of selected functionals.

Table S7: linear fit parameters for Figure S2.

| functional | Intercept±SE [kHz] | Slope±SE | $R^2$ |
|---|---|---|---|
| **PBE** | 31 ± 19 | 0.89 ± 0.05 | 0.98 |
| **PBE+D3** | 20 ± 13 | 0.93 ± 0.04 | 0.99 |
| **revPBE+D3** | 1 ± 19 | 0.87 ± 0.05 | 0.98 |
| **PBEsol+D3** | 40 ± 25 | 0.99 ± 0.07 | 0.97 |
| **rVV10** | 63 ± 27 | 1.01 ± 0.08 | 0.97 |
| **rev-vdW-DF2** | 54 ± 19 | 1.00 ± 0.05 | 0.99 |
| **vdW-DF-c6** | 62 ± 22 | 1.00 ± 0.06 | 0.98 |
| **rPBE+D2* Zhang** | 29 ± 18 | 0.89 ± 0.10 | 0.96 |

## ⁸⁷Rb Quadrupolar coupling constants

The quadrupolar coupling constant of $^{133}$Cs is small due its low quadrupolar moment (-0.00343 b[24]). This raised the question whether the differences between Cq values obtained by various functionals fall within the error range. To check this, we performed similar calculations using a nucleus with a much larger quadrupolar moment (0.1335 b[24]) – $^{87}$Rb. The results (Figure S3) show the same trend observed in $^{133}$Cs repeats itself in $^{87}$Rb. We therefore conclude that the differences between Cq of various functionals reflect inherent differences between these functionals.

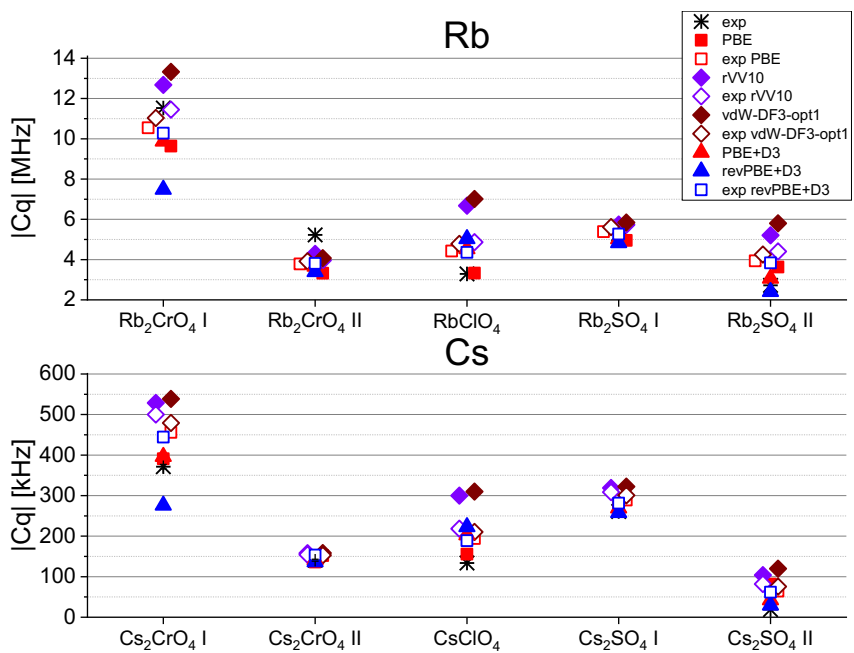

*Figure S3: Quadrupolar coupling constants of Rb salts, compared with Cs salts. The experimental $^{87}$Rb data are taken from previously reported studies[25–27].*

## Chemical Shifts

Table S8 – S12 specify the experimental and computational $^{133}$Cs chemical shift parameters of the materials discussed in the main text.

To generate consistent experimental values, we re-referenced the results of several studies to fit a uniform scale, where solid CsCl resonates at 223.2 ppm, as commonly used in solid-state NMR studies. Since some studies also used 0.5M or 1.0M solutions of CsCl as a reference[28] and some studies were performed at static conditions, we also acquired our own data at room temperature using a 14.1T magnet and 5 kHz spinning. We measured solid samples of CsCl, CsI, Cs$_2$CrO$_4$, and Cs$_2$SO$_4$. We also performed experiments on 5 solutions of CsCl at different concentrations at the range of 0.1-1.0 [M] and obtained a linear fit where $\delta_{iso}(^{133}\text{Cs, ppm}) = 11.3(0.2)\left[\frac{M}{\text{ppm}}\right] - 6.1(0.2)[\text{ppm}]$. This result was obtained by referencing all data to solid CsCl at 223.2 ppm and is consistent with infinite dilution values reported by Haase[29] though with some deviations for his reported value of 0.5M CsCl.

Our experimental results are shown in the top of Figure S4. To demonstrate the relation between solid CsCl and a 0.1M solution of CsNO$_3$ in D$_2$O, as recommended by IUPAC, we measured the two samples at static conditions and at two temperatures (bottom of Figure S4). While the solid sample is not affected by the temperature, the solution shows a shift of 1.9 ppm over 15°C, suggesting a strong temperature dependence.

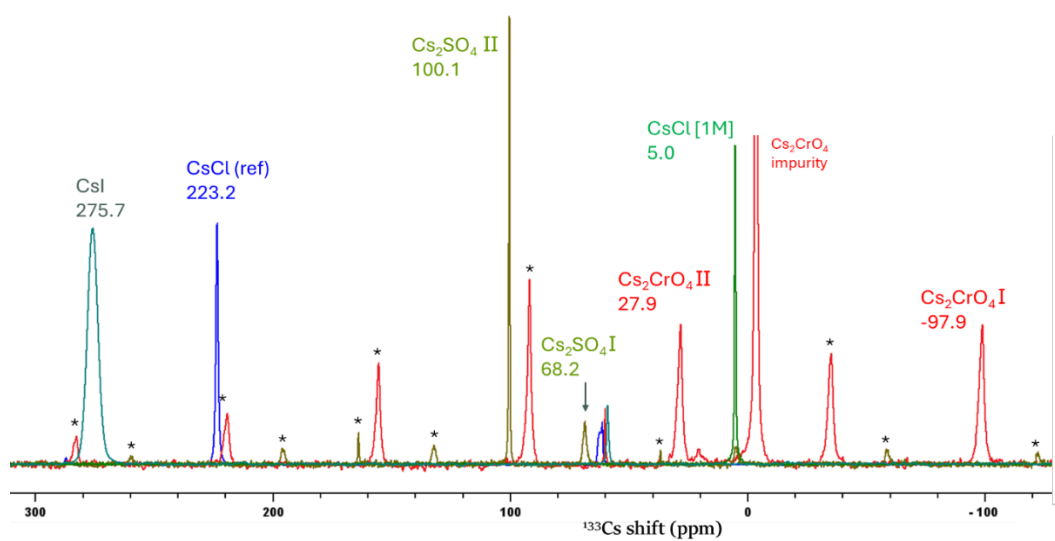

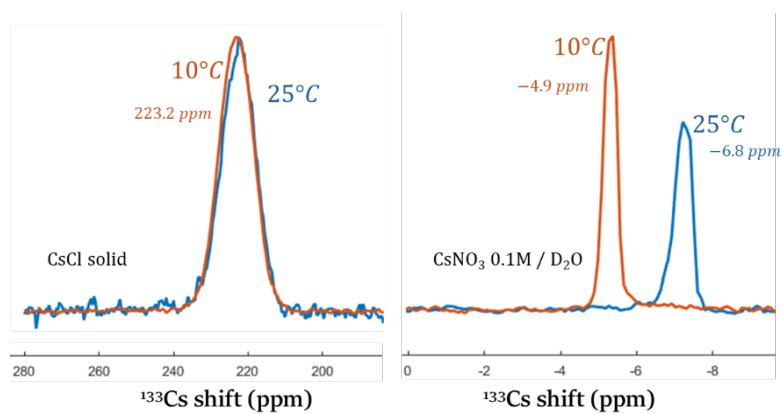

Figure S4: Top: Overlay of $^{133}$Cs 5 kHz magic-angle spinning solid-state NMR spectra of Cs salts measured on a 14.1T magnet. All data were referenced to solid CsCl at 223.2 ppm. Powder samples were used as purchased and contain some minor contaminants. The Cs$_2$CrO$_4$ sample has a large contaminant at ~0 ppm but the two sites are clearly resolved. Spinning sidebands, marked by asterisks, were determined by collecting additional experiments at 6.66 and 7.50 kHz. Bottom: Static spectra of powdered CsCl (left) and a 0.1M solution of CsNO$_3$ in D$_2$O (right, IUPAC recommendation) at 10°C and 25°C. The powder of CsNO$_3$ was initially dried in the oven since it is highly hygroscopic.

*Table S8: Experimental chemical shifts [ppm] of the materials examined in this work.*

|  | Experimental | | | |
|---|---|---|---|---|
|  | $\delta_{iso}$ | $\Delta\delta = \delta_{zz} - (\delta_{xx} + \delta_{yy})/2$ | $\eta = (\delta_{yy} - \delta_{xx})/(\delta_{zz} - \delta_{iso})$ | ref |
| **CsF** | 181.3* | 0 | 0 | 28 |
| **CsCl** | 223.2 | 0 | 0 | 30 |
| **CsBr** | 256.0* | 0 | 0 | 28 |
|  | 258.2 | 0 | 0 | 30 |
| **CsI** | 276.1* | 0 | 0 | 28 |
|  | 275.7 | 0 | 0 | This work |
| **Cs$_2$CrO$_4$ I** | -98.3^ | -328.5 | 0.04 | 18 |
|  | -98.3 | -331.5 | 0.06 | 16 |
|  | -100 | -333 | 0.04 | 17 |
|  | -97.9 |  |  | This work |
| **Cs$_2$CrO$_4$ II** | 28.7^ | 246 | 0.3 | 18 |
|  | 28.2 | 243 | 0.31 | 16 |
|  | 27.0 | 244.5 | 0.26 | 17 |
|  | 27.9 |  |  | This work |
| **CsClO$_4$** | 2.7 | 34.35 | 0.32 | 16 |
| **Cs$_2$SO$_4$ I** | 68.6 | -22.45 | 0.14 | 16 |
|  | 68.2 |  |  | This work |
| **Cs$_2$SO$_4$ II** | 100.3 | -46.5 | 0.49 | 16 |
|  | 100.1 |  |  | This work |
| **CsVO$_3$** | -32.0 | -135 | 0.44 | 16 |
| **CsGeCl$_3$** | 36.9 | 10 | 0 | 7 |
|  | 33.7 | 10 | 0 |  |
| **CsGeBr$_3$** | 48.2 | -9 | 1 | 7 |
|  | 47.5 | 16 | 0 |  |
| **CsGeI$_3$** | 41.8 | NA | NA | 7 |
|  | 39.2 |  |  |  |
| **CsCd(SCN)$_3$** | 54.7^ | 83.8 | 0.38 | 19 |
| **CsBPh$_4$** | -272* | 36.5 | 0 | 20 |
| **[Cs$^+$(Cryptand[2.2.2])]I$^-$** | 225 | 39 | 0.92 | 31 |
| **CsB$_3$O$_5$** | 70.9* | 120 | NA | 21 |
| **CsSc$_3$F$_{10}$** | -1.8^ | NA | NA | 12 |
| **CsPbI$_3$, hexagonal** | 245* | NA | NA | 32 |
| **CsPbI$_3$, tetragonal** | 119.1* | NA | NA | 32 |

\* Originally referenced to 1.0 M CsCl=0 ppm. Our measurements show that with respect to solid CsCl at 223.2 ppm, the shift is +5.0 ppm.

^ Originally referenced to 0.5 M CsCl=0 ppm. Our measurements show that with respect to solid CsCl at 223.2 ppm, the shift is -0.6 ppm.

*Table S9: Calculated Chemical shifts [ppm] – PBE functionals*

|  | PBE | | | Exp PBE | | | PBE + D2 Zhang | | | PBE+D3 | | | PBE+XDM | | |
|---|---|---|---|---|---|---|---|---|---|---|---|---|---|---|---|
|  | $\delta_{iso}$ | $\Delta\delta$ | $\eta$ | $\delta_{iso}$ | $\Delta\delta$ | $\eta$ | $\delta_{iso}$ | $\Delta\delta$ | $\eta$ | $\delta_{iso}$ | $\Delta\delta$ | $\eta$ | $\delta_{iso}$ | $\Delta\delta$ | $\eta$ |
| **CsF** | 198.8 | 0 | 0 | 200.2 | 0 | 0 | 196.3 | 0 | 0 | 209.3 | 0 | 0 | 226.3 | 0 | 0 |
| **CsCl** | 230.7 | 0 | 0 | 226.5 | 0 | 0 | 216.4 | 0 | 0 | 226.3 | 0 | 0 | 241.6 | 0 | 0 |
| **CsBr** | 261.6 | 0 | 0 | 268.3 | 0 | 0 | 251.9 | 0 | 0 | 254.7 | 0 | 0 | 259.1 | 0 | 0 |
| **CsI** | 270.5 | 0 | 0 | 276.2 | 0 | 0 | 291.7 | 0 | 0 | 266.0 | 0 | 0 | 273.1 | 0 | 0 |
| **Cs$_2$CrO$_4$ I** | -69.4 | -277.0 | 0.16 | -69.5 | -309.3 | 0.28 | -85.7 | -347.0 | 0.42 | -84.7 | -276.5 | 0.03 | -41.2 | -453.6 | 0.44 |
| **Cs$_2$CrO$_4$ II** | 26.8 | 356.3 | 0.01 | 37.2 | 378.0 | 0.02 | 32.1 | 389.2 | 0.07 | 25.5 | 350.3 | 0.03 | 56.1 | 541.5 | 0.05 |
| **CsClO$_4$** | 5.4 | 43.9 | 0.38 | 14.1 | 50.2 | 0.90 | -0.3 | 63.1 | 0.19 | 15.5 | 52.9 | 0.25 | 7.8 | 81.1 | 0.56 |
| **Cs$_2$SO$_4$ I** | 74.1 | 38.1 | 0.91 | 69.3 | -47.0 | 0.16 | 57.3 | -46.0 | 0.48 | 72.7 | -38.5 | 0.79 | 59.7 | -78.7 | 0.20 |
| **Cs$_2$SO$_4$ II** | 103.8 | 19.8 | 0.62 | 102.5 | 17.2 | 0.16 | 90.2 | -10.5 | 0.58 | 103.9 | -11.3 | 0.50 | 96.5 | -30.4 | 0.45 |
| **CsVO$_3$** | -34.7 | -165.9 | 0.30 | -13.6 | -172.0 | 0.46 | -19.6 | -186.2 | 0.58 | -36.1 | -159.7 | 0.45 | -29.1 | -190.5 | 0.28 |
| **CsGeCl$_3$** | 9.8 | -19.9 | 0 | -9.6 | -16.9 | 0.00 | 26.6 | -9.4 | 0.00 | 12.2 | -12.5 | 0.00 | 3.2 | -9.9 | 0.00 |
| **CsGeBr$_3$** | 29.8 | -24.8 | 0 | 13.7 | -23.5 | 0.00 | 42.0 | -11.0 | 0.00 | 38.4 | -17.5 | 0.00 | -16.5 | -22.5 | 0.00 |
| **CsGeI$_3$** | 18.9 | -12.9 | 0 | 10.9 | -14.9 | 0.00 | 27.2 | -9.5 | 0.00 | 22.5 | -9.5 | 0.00 | -10.4 | -22.2 | 0.00 |
| **CsCd(SCN)$_3$** | 10.8 | 120.8 | 0.40 | | | | | | | 32.7 | 111.4 | 0.38 | | | |
| **CsBPh$_4$** | -223.7 | 48.5 | 0.00 | | | | | | | -250.8 | 55.1 | 0.00 | | | |
| **[Cs$^+$(Cryptand[2.2.2])]I$^-$** | 254.3 | -22.6 | 0.92 | | | | | | | 264.9 | -9.7 | 0.33 | | | |
| **CsB$_3$O$_5$** | 70.2 | 106.7 | 0.67 | | | | | | | 69.5 | 95.6 | 0.88 | | | |
| **CsSc$_3$F$_{10}$** | 42.8 | -116.1 | 0.41 | | | | | | | 29.5 | -118.2 | 0.41 | | | |
| **CsPbI$_3$, hexagonal** | 228.7 | | | | | | | | | 265.8 | | | | | |
| **CsPbI$_3$, tetragonal** | 184.1 | | | | | | | | | 127.2 | | | | | |

*Table S10: Calculated Chemical shifts [ppm] – rPBE functionals*

| | rPBE + D2 Zhang | | | rPBE +D2* | | | rPBE +D2* Zhang | | | exp rPBE | | |
|---|---|---|---|---|---|---|---|---|---|---|---|---|
| | $\delta_{iso}$ | $\Delta\delta$ | $\eta$ | $\delta_{iso}$ | $\Delta\delta$ | $\eta$ | $\delta_{iso}$ | $\Delta\delta$ | $\eta$ | $\delta_{iso}$ | $\Delta\delta$ | $\eta$ |
| **CsF** | 178.3 | 0.0 | 0.00 | 323.8 | 0.0 | 0.00 | 199.0 | 0.0 | 0.00 | 198.8 | 0.0 | 0.00 |
| **CsCl** | 210.6 | 0.0 | 0.00 | 234.3 | 0.0 | 0.00 | 210.0 | 0.0 | 0.00 | 221.0 | 0.0 | 0.00 |
| **CsBr** | 282.8 | 0.0 | 0.00 | 247.9 | 0.0 | 0.00 | 277.4 | 0.0 | 0.00 | 268.5 | 0.0 | 0.00 |
| **CsI** | 291.6 | 0.0 | 0.00 | 238.5 | 0.0 | 0.00 | 280.4 | 0.0 | 0.00 | 275.4 | 0.0 | 0.00 |
| **$Cs_2CrO_4$ I** | -60.0 | -273.1 | 0.31 | -21.8 | -553.1 | 0.36 | -58.8 | -262.4 | 0.23 | -80.6 | -281.6 | 0.27 |
| **$Cs_2CrO_4$ II** | 38.4 | 313.6 | 0.06 | 91.5 | 331.9 | 0.35 | 30.5 | 331.6 | 0.01 | 22.2 | 351.5 | 0.01 |
| **$CsClO_4$** | 0.6 | 48.3 | 0.12 | -29.5 | 71.2 | 0.40 | -7.4 | 40.7 | 0.42 | 13.2 | 50.2 | 0.90 |
| **$Cs_2SO_4$ I** | 69.2 | 33.8 | 0.89 | 88.1 | -81.3 | 0.23 | 74.7 | -37.2 | 0.81 | 78.8 | -42.4 | 0.24 |
| **$Cs_2SO_4$ II** | 93.4 | 8.4 | 0.88 | 99.8 | -60.7 | 0.97 | 102.1 | 16.7 | 0.56 | 110.2 | 14.8 | 0.30 |
| **$CsVO_3$** | -20.6 | -151.5 | 0.41 | -13.4 | 209.1 | 0.99 | -22.6 | -152.6 | 0.34 | 2.1 | -158.8 | 0.47 |
| **$CsGeCl_3$** | -11.0 | -21.4 | 0.00 | -49.8 | -12.0 | 0.00 | -14.4 | -19.6 | 0.00 | -10.6 | -15.2 | 0.00 |
| **$CsGeBr_3$** | 22.5 | -25.3 | 0.00 | -38.8 | -13.6 | 0.00 | 18.8 | -24.9 | 0.00 | 14.0 | -22.8 | 0.00 |
| **$CsGeI_3$** | 30.3 | -10.0 | 0.00 | -44.4 | -17.5 | 0.00 | 36.5 | -14.6 | 0.00 | 13.2 | -15.7 | 0.00 |

*Table S11: Calculated Chemical shifts [ppm] – additional semilocal functionals*

| | revPBE + D3 | | | exp revPBE | | | PBEsol + D2 Zhang | | | PBEsol + D3 | | | exp PBEsol | | | B86bPBE + XDM | | | exp B86bPBE | | |
|---|---|---|---|---|---|---|---|---|---|---|---|---|---|---|---|---|---|---|---|---|---|
| | $\delta_{iso}$ | $\Delta\delta$ | $\eta$ | $\delta_{iso}$ | $\Delta\delta$ | $\eta$ | $\delta_{iso}$ | $\Delta\delta$ | $\eta$ | $\delta_{iso}$ | $\Delta\delta$ | $\eta$ | $\delta_{iso}$ | $\Delta\delta$ | $\eta$ | $\delta_{iso}$ | $\Delta\delta$ | $\eta$ | $\delta_{iso}$ | $\Delta\delta$ | $\eta$ |
| CsF | 195.1 | 0.0 | 0.00 | 196.9 | 0.0 | 0.00 | 220.1 | 0.0 | 0.00 | 201.1 | 0.0 | 0.00 | 201.7 | 0.0 | 0.00 | 198.3 | 0.0 | 0.00 | 200.1 | 0.0 | 0.00 |
| CsCl | 235.9 | 0.0 | 0.00 | 230.6 | 0.0 | 0.00 | 217.3 | 0.0 | 0.00 | 223.9 | 0.0 | 0.00 | 238.5 | 0.0 | 0.00 | 238.0 | 0.0 | 0.00 | 224.9 | 0.0 | 0.00 |
| CsBr | 261.8 | 0.0 | 0.00 | 268.6 | 0.0 | 0.00 | 250.9 | 0.0 | 0.00 | 254.5 | 0.0 | 0.00 | 262.7 | 0.0 | 0.00 | 265.5 | 0.0 | 0.00 | 268.3 | 0.0 | 0.00 |
| CsI | 250.0 | 0.0 | 0.00 | 277.3 | 0.0 | 0.00 | 279.9 | 0.0 | 0.00 | 275.6 | 0.0 | 0.00 | 270.3 | 0.0 | 0.00 | 278.0 | 0.0 | 0.00 | 275.1 | 0.0 | 0.00 |
| $Cs_2CrO_4$ I | -101.7 | -219.6 | 0.16 | -69.5 | -288.2 | 0.26 | -88.6 | -420.8 | 0.46 | -85.9 | -353.7 | 0.28 | -66.0 | -321.8 | 0.31 | -69.1 | -368.7 | 0.32 | -80.0 | -305.4 | 0.26 |
| $Cs_2CrO_4$ II | 19.8 | 265.8 | 0.08 | 34.4 | 352.6 | 0.00 | 37.7 | 479.2 | 0.07 | 24.7 | 436.6 | 0.03 | 37.5 | 406.3 | 0.04 | 27.6 | 446.4 | 0.04 | 26.2 | 378.9 | 0.00 |
| $CsClO_4$ | 14.6 | 47.3 | 0.05 | 13.6 | 47.7 | 0.89 | -1.8 | 73.4 | 0.48 | 0.9 | 62.6 | 0.50 | 8.6 | 51.5 | 0.99 | 8.9 | 72.9 | 0.43 | 26.3 | 50.2 | 0.90 |
| $Cs_2SO_4$ I | 74.6 | 32.5 | 0.67 | 72.0 | -43.0 | 0.23 | 52.2 | -64.6 | 0.06 | 61.0 | -54.2 | 0.11 | 71.4 | -50.9 | 0.10 | 55.8 | -78.7 | 0.20 | 73.1 | -47.0 | 0.16 |
| $Cs_2SO_4$ II | 101.1 | -5.2 | 0.77 | 104.2 | 15.1 | 0.32 | 92.3 | -13.3 | 0.85 | 97.0 | -14.1 | 0.86 | 103.9 | 18.8 | 0.04 | 91.1 | -30.4 | 0.45 | 106.2 | 17.2 | 0.16 |
| $CsVO_3$ | -36.1 | -128.1 | 0.52 | -1.7 | -158.5 | 0.46 | -12.0 | -214.8 | 0.70 | -28.1 | -195.4 | 0.59 | -18.3 | -183.9 | 0.50 | -3.0 | -189.1 | 0.65 | -0.8 | -168.9 | 0.44 |
| $CsGeCl_3$ | 21.4 | -11.3 | 0.00 | -12.0 | -16.4 | 0.00 | 46.0 | -8.5 | 0.00 | 33.4 | -8.4 | 0.00 | -5.6 | -17.6 | 0.00 | 18.0 | -5.8 | 0.00 | -12.3 | -16.0 | 0.00 |
| $CsGeBr_3$ | 61.6 | -16.1 | 0.00 | 11.3 | -23.2 | 0.00 | 6.3 | -21.1 | 0.00 | 46.9 | -13.0 | 0.00 | 15.5 | -24.7 | 0.00 | 16.9 | -10.9 | 0.00 | 15.0 | -23.4 | 0.00 |
| $CsGeI_3$ | 28.1 | -10.8 | 0.00 | 0.4 | -14.5 | 0.00 | 25.8 | -19.9 | 0.00 | 21.3 | -11.1 | 0.00 | 6.1 | -14.0 | 0.00 | 0.2 | -12.7 | 0.00 | 4.1 | -13.7 | 0.00 |
| $CsCd(SCN)_3$ | 38.3 | 84.6 | 0.62 | | | | | | | 47.3 | 137.5 | 0.46 | | | | | | | | | |
| $CsBPh_4$ | -320.8 | -3.6 | 0.00 | | | | | | | -233.6 | 74.0 | 0.00 | | | | | | | | | |
| $[Cs^+(C222)]I^-$ | 349.6 | 20.0 | 0.39 | | | | | | | 204.3 | 16.6 | 0.42 | | | | | | | | | |
| $CsB_3O_5$ | 57.9 | 74.0 | 0.92 | | | | | | | 52.8 | 113.6 | 0.90 | | | | | | | | | |
| $CsSc_3F_{10}$ | 48.5 | -99.8 | 0.41 | | | | | | | -4.1 | -140.7 | 0.44 | | | | | | | | | |
| $CsPbI_3$, hexagonal | 283.6 | | | | | | | | | 272.4 | | | | | | | | | | | |
| $CsPbI_3$, tetragonal | -37.7 | | | | | | | | | 130.0 | | | | | | | | | | | |

*Table S12: Calculated Chemical shifts [ppm] – non-local functionals*

| | rVV10 | | | exp rVV10 | | | vdW-DF3-opt1 | | | exp vdW-DF3-opt1 | | | vdW-DF-C6 | | | exp vdW-DF-C6 | | | rev-vdW-DF2 | | | exp rev-vdW-DF2 | | |
|---|---|---|---|---|---|---|---|---|---|---|---|---|---|---|---|---|---|---|---|---|---|---|---|---|
| | $\delta_{iso}$ | $\Delta\delta$ | $\eta$ | $\delta_{iso}$ | $\Delta\delta$ | $\eta$ | $\delta_{iso}$ | $\Delta\delta$ | $\eta$ | $\delta_{iso}$ | $\Delta\delta$ | $\eta$ | $\delta_{iso}$ | $\Delta\delta$ | $\eta$ | $\delta_{iso}$ | $\Delta\delta$ | $\eta$ | $\delta_{iso}$ | $\Delta\delta$ | $\eta$ | $\delta_{iso}$ | $\Delta\delta$ | $\eta$ |
| CsF | 190.5 | 0.0 | 0.00 | 194.6 | 0.0 | 0.00 | 196.6 | 0.0 | 0.00 | 191.1 | 0.0 | 0.00 | 188.9 | 0.0 | 0.00 | 210.6 | 0.0 | 0.00 | 185.8 | 0.0 | 0.00 | 193.4 | 0.0 | 0.00 |
| CsCl | 223.6 | 0.0 | 0.00 | 227.4 | 0.0 | 0.00 | 240.0 | 0.0 | 0.00 | 226.2 | 0.0 | 0.00 | 233.1 | 0.0 | 0.00 | 242.8 | 0.0 | 0.00 | 231.4 | 0.0 | 0.00 | 223.9 | 0.0 | 0.00 |
| CsBr | 254.5 | 0.0 | 0.00 | 266.0 | 0.0 | 0.00 | 226.2 | 0.0 | 0.00 | 267.6 | 0.0 | 0.00 | 241.2 | 0.0 | 0.00 | 281.2 | 0.0 | 0.00 | 251.7 | 0.0 | 0.00 | 263.4 | 0.0 | 0.00 |
| CsI | 279.2 | 0.0 | 0.00 | 276.0 | 0.0 | 0.00 | 283.4 | 0.0 | 0.00 | 273.2 | 0.0 | 0.00 | 280.5 | 0.0 | 0.00 | 224.0 | 0.0 | 0.00 | 275.4 | 0.0 | 0.00 | 273.2 | 0.0 | 0.00 |
| $Cs_2CrO_4$ I | -99.4 | -352.5 | 0.30 | -74.1 | -341.5 | 0.30 | -101.8 | -359.9 | 0.30 | -83.8 | -325.5 | 0.30 | -100.4 | -339.2 | 0.30 | -84.5 | -327.0 | 0.30 | -95.7 | -331.7 | 0.30 | -86.9 | -325.8 | 0.30 |
| $Cs_2CrO_4$ II | 21.9 | 433.4 | 0.00 | 37.9 | 426.4 | 0.00 | 19.3 | 441.2 | 0.00 | 26.6 | 410.0 | 0.00 | 19.1 | 419.1 | 0.00 | 34.9 | 405.7 | 0.00 | 19.6 | 411.9 | 0.00 | 22.1 | 404.6 | 0.00 |
| $CsClO_4$ | 39.1 | 74.2 | 0.23 | 14.7 | 54.7 | 0.87 | 30.4 | 71.5 | 0.36 | 17.8 | 52.1 | 0.87 | 29.8 | 68.6 | 0.23 | -0.2 | 50.2 | 0.90 | 25.9 | 64.4 | 0.29 | 26.4 | 53.0 | 0.84 |
| $Cs_2SO_4$ I | 68.9 | -51.2 | 0.38 | 64.3 | -54.6 | 0.04 | 68.6 | -54.8 | 0.21 | 77.4 | -51.4 | 0.09 | 70.6 | -51.0 | 0.28 | 83.6 | -52.2 | 0.04 | 71.9 | -49.4 | 0.32 | 74.1 | -51.6 | 0.04 |
| $Cs_2SO_4$ II | 107.3 | -18.3 | 0.52 | 98.5 | 21.0 | 0.09 | 111.0 | -18.0 | 0.87 | 112.0 | 19.8 | 0.09 | 110.0 | -16.0 | 0.95 | 120.4 | 19.7 | 0.10 | 109.3 | -15.3 | 1.00 | 107.8 | 19.6 | 0.07 |
| $CsVO_3$ | -21.1 | -189.9 | 0.51 | -25.4 | -192.5 | 0.45 | -26.1 | -193.6 | 0.58 | -5.4 | -179.1 | 0.43 | -25.3 | -183.8 | 0.51 | -26.0 | -181.7 | 0.47 | -28.2 | -181.8 | 0.51 | -21.8 | -182.5 | 0.49 |
| $CsGeCl_3$ | 28.0 | -7.2 | 0.00 | -2.5 | -17.3 | 0.00 | 35.0 | -9.5 | 0.00 | -3.9 | -17.5 | 0.00 | 30.0 | -13.1 | 0.00 | -3.3 | -16.6 | 0.00 | 32.2 | -13.2 | 0.00 | 6.2 | -18.3 | 0.00 |
| $CsGeBr_3$ | 29.9 | -11.9 | 0.00 | 27.3 | -24.2 | 0.00 | 36.9 | -12.7 | 0.00 | 18.3 | -24.2 | 0.00 | 36.1 | -12.4 | 0.00 | 26.1 | -23.8 | 0.00 | 35.4 | -14.0 | 0.00 | 26.7 | -23.9 | 0.00 |
| $CsGeI_3$ | 3.8 | -3.3 | 0.00 | 21.6 | -11.8 | 0.00 | 6.6 | -8.5 | 0.00 | 8.9 | -12.9 | 0.00 | 12.4 | -6.3 | 0.00 | 16.6 | -12.4 | 0.00 | 11.5 | -6.6 | 0.00 | 17.5 | -12.5 | 0.00 |
| $CsCd(SCN)_3$ | 66.4 | 137.4 | 0.85 | | | | | | | | | | 68.6 | 138.0 | 0.39 | | | | 63.7 | 136.3 | 0.39 | | | |
| $CsBPh_4$ | -252.5 | 67.8 | 0.00 | | | | | | | | | | -244.1 | 73.7 | 0.00 | | | | -237.3 | 73.0 | 0.00 | | | |
| [Cs⁺(C222)]I⁻ | 198.7 | 30.4 | 0.04 | | | | | | | | | | 223.6 | 22.7 | 0.09 | | | | 228.7 | 19.2 | 0.11 | | | |
| $CsB_3O_5$ | 77.7 | 118.1 | 0.96 | | | | | | | | | | 71.7 | 118.0 | 0.84 | | | | 69.5 | 110.5 | 0.81 | | | |
| $CsSc_3F_{10}$ | 12.8 | -122.9 | 0.05 | | | | | | | | | | -9.1 | -129.8 | 0.38 | | | | -1.6 | -130.4 | 0.41 | | | |
| $CsPbI_3$, hexagonal | 274.0 | | | | | | | | | | | | 271.7 | | | | | | 257.6 | | | | | |
| $CsPbI_3$, tetragonal | 238.7 | | | | | | | | | | | | 221.7 | | | | | | 219.8 | | | | | |

## Chemical Shift Anisotropy

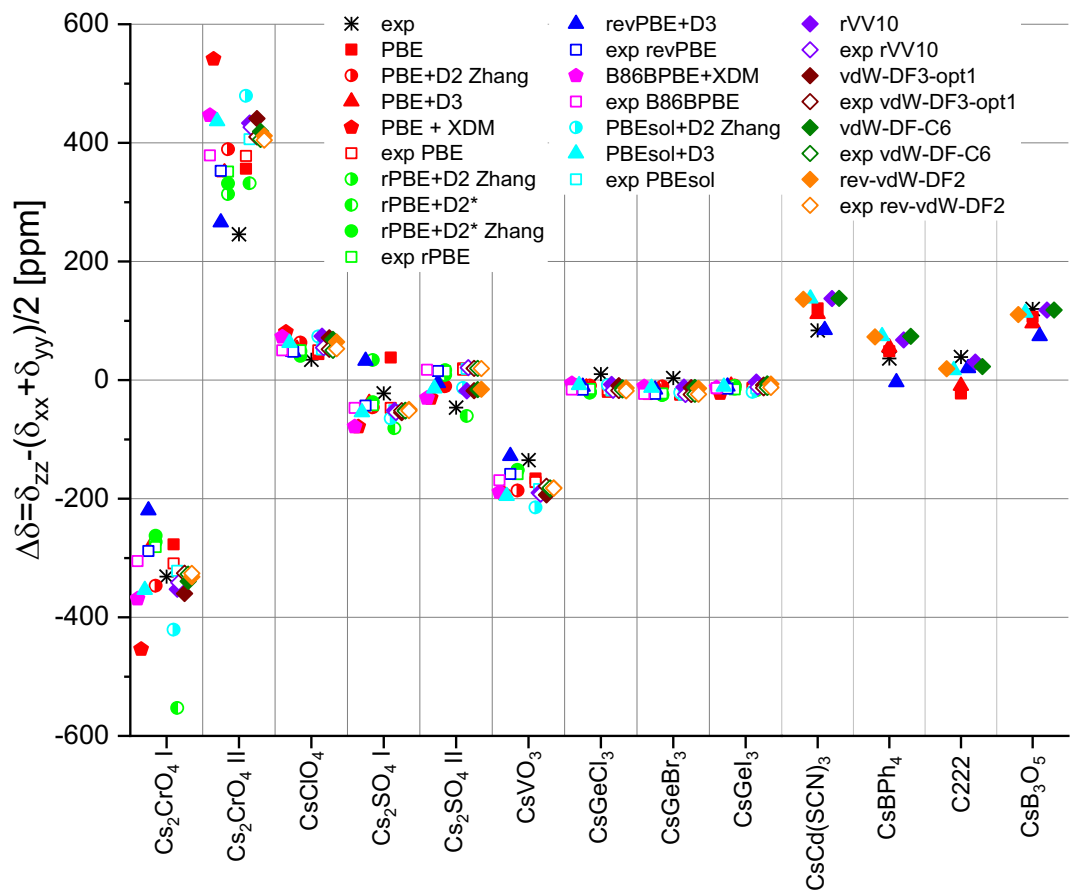

*Figure S5: Calculated vs experimental chemical shift anisotropies.*

## Fully relativistic geometry optimization

The calculations in this work were performed using scalar-relativistic pseudopotentials, as GIPAW currently does not support non-collinear calculations and geometry calculations require heavy computational resources. In order to ascertain the validity of this approach, we carried out one geometry optimization using fully relativistic pseudopotentials. Then, the NMR parameters were calculated using scalar-relativistic PPs. The functional used was PBE. The results (Table S13) show that the use of the scalar-relativistic approximation for the geometry optimization is justified as the errors are below our best MAEs. It is however still possible that spin orbit coupling effects may have a sizable contribution to the NMR parameters and our MAEs may be further reduced using this approach in future studies.

*Table S13: results of scalar-relativistic vs fully relativistic calculations on $Cs_2CrO_4$.*

|  | Unit cell volume [Å³] | | | Optimization wall time on 32 cores [d] | | |Cq| [kHz] | | | $\delta_{iso}$ [ppm] | | |
|---|---|---|---|---|---|---|---|---|---|---|---|
|  | exp | Scalar relativistic | Fully relativistic | Scalar relativistic | Fully relativistic | exp | Scalar relativistic | Fully relativistic | exp | Scalar relativistic | Fully relativistic |
| site I | 594.61 | 630.10 | 634.70 | 3.5 | 8.5 | 365[16] 376[17] 373[18] | 390.8 | 388.6 | -98.3[18] -100[17] | -69.4 | -64.2 |
| site II |  |  |  |  |  | 142[16] 138[17] 142[18] | -136.1 | -130.1 | 28.7[18] 28.2[16] 27.0[17] | 26.8 | 27.4 |

## Comparison with fully relativistic chemical shift calculations

Very few examples of fully relativistic $^{133}$Cs chemical shift DFT calculations are available in the literature. One of them[33] does not include any comparison with experimental data, preventing an assessment of the importance of SOC incorporation. The other[32] provides both the experimental chemical shift values and fully-relativistic computational results, allowing such an assessment.

The calculations in ref [32] were carried out by fully-relativistic PBE+D3 optimization, followed by the generation of local clusters of a central cation surrounded by a PbI$_3$ cage representing the asymmetric unit of the periodic crystal structure, which was used for a fully relativistic BP86+D3 calculation of chemical shielding. Our methods therefore differ from those not only by the inclusion of SOC but also by the use of the cluster method and of the BP86 functional, which were not tested here. The results (Figure S6) show that the fully relativistic calculation sometimes outperforms the scalar-relativistic one to an extent which is highly system-dependent. This phenomenon is mediated not only by Cs but obviously also by Pb. In the absence of other systems for which both experimental $^{133}$Cs chemical shift values and fully-relativistic computational ones are available, it is hard to draw any further conclusions.

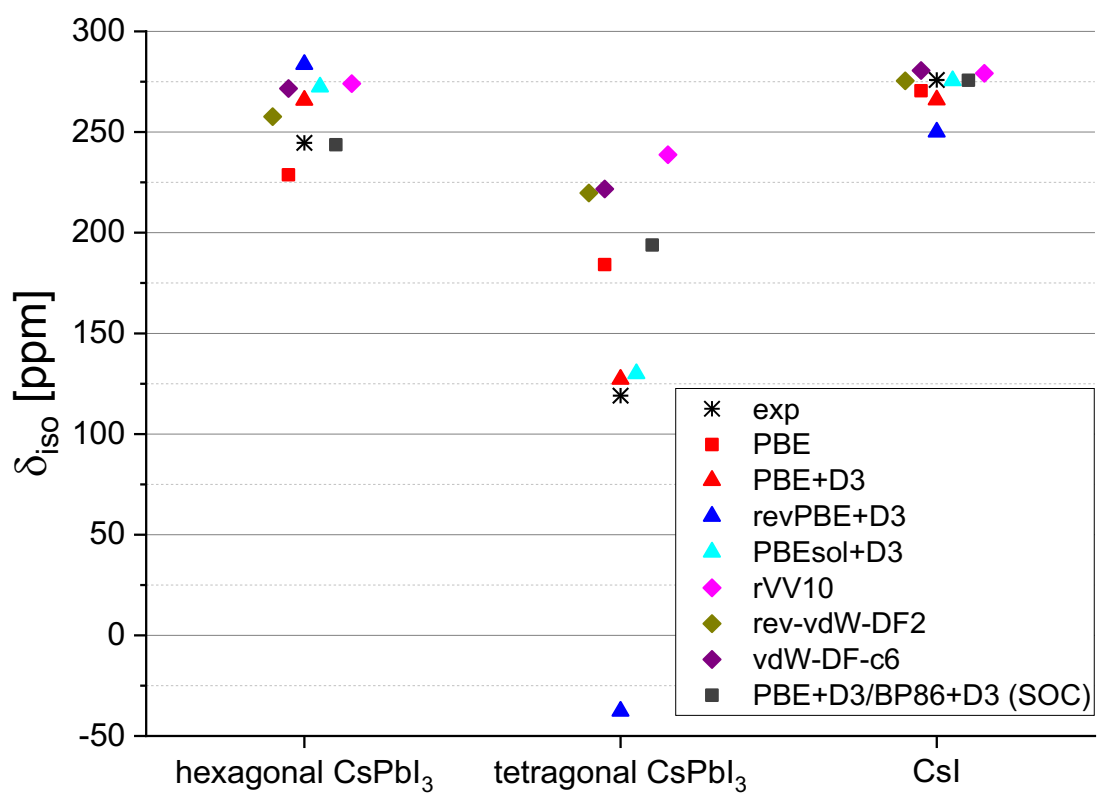

*Figure S6: Comparison of chemical shift values calculated with and without SOC. The values with SOC, as well as the experimental values, are taken from[32].*

## References

(1) Davey, W. P. Precision Measurements of Crystals of the Alkali Halides. *Phys. Rev.* **1923**, *21* (2), 143.
(2) Wyckoff, R. W. *Crystal Structures*, 2nd ed.; Interscience: New York, 1963; Vol. 5.
(3) Morris, A.; Kennard, C.; Moore, F.; Smith, G.; Berendsen, H. J. C. Cesium Chromate, CrCs2O4 (Neutron). *Cryst. Struct. Commun.* **1981**, *10* (3), 529–532.
(4) Granzin, J. Refinement of the Crystal Structures of RbClO4 and CsClO4. *Z. Für Krist.-Cryst. Mater.* **1988**, *184* (1–4), 157–160.
(5) Weber, H.; Schulz, M.; Schmitz, S.; Granzin, J.; Siegert, H. Determination and Structural Application of Anisotropic Bond Polarisabilities in Complex Crystals. *J. Phys. Condens. Matter* **1989**, *1* (44), 8543.
(6) Hawthorne, F.; Calvo, C. The Crystal Chemistry of the M+ VO3 (M+= Li, Na, K, NH4, Tl, Rb, and Cs) Pyroxenes. *J. Solid State Chem.* **1977**, *22* (2), 157–170.
(7) Hooper, R. W.; Ni, C.; Tkachuk, D. G.; He, Y.; Terskikh, V. V.; Veinot, J. G.; Michaelis, V. K. Exploring Structural Nuances in Germanium Halide Perovskites Using Solid-State 73Ge and 133Cs NMR Spectroscopy. *J. Phys. Chem. Lett.* **2022**, *13* (7), 1687–1696.
(8) Thiele, G.; Messer, D. S-Thiocyanato-und N-Isothiocyanato-Bindungsisomerie in Den Kristallstrukturen von RbCd (SCN) 3 Und CsCd (SCN) 3. *Z. Für Anorg. Allg. Chem.* **1980**, *464* (1), 255–267.
(9) Behrens, U.; Hoffmann, F.; Olbrich, F. Solid-State Structures of Base-Free Lithium and Sodium Tetraphenylborates at Room and Low Temperature: Comparison with the Higher Homologues MB (C6H5) 4 (M= K, Rb, Cs). *Organometallics* **2012**, *31* (3), 905–913.
(10) Moon, C. J.; Park, J.; Im, H.; Ryu, H.; Choi, M. Y.; Kim, T. H.; Kim, J. Chemical Shift and Second-Order Quadrupolar Effects in the Solid-State 133Cs NMR Spectra of [Cs+ (Cryptand [2.2. 2])] X (X= I−, SCN−· H2O). *Bull. Korean Chem. Soc.* **2020**, *41* (7), 702–708.
(11) Kaneko, S.; Tokuda, Y.; Takahashi, Y.; Masai, H.; Ueda, Y. Structural Analysis of Mixed Alkali Borosilicate Glasses Containing Cs+ and Na+ Using Strong Magnetic Field Magic Angle Spinning Nuclear Magnetic Resonance. *J. Asian Ceram. Soc.* **2017**, *5* (1), 7–12.
(12) Rakhmatullin, A.; Allix, M.; Polovov, I. B.; Maltsev, D.; Chukin, A. V.; Bakirov, R.; Bessada, C. Combining Solid State NMR, Powder X-Ray Diffraction, and DFT Calculations for CsSc3F10 Structure Determination. *J. Alloys Compd.* **2019**, *787*, 1349–1355.
(13) Ying, Y.; Luo, X.; Huang, H. Pressure-Induced Topological Nontrivial Phase and Tunable Optical Properties in All-Inorganic Halide Perovskites. *J. Phys. Chem. C* **2018**, *122* (31), 17718–17725.
(14) Trots, D.; Myagkota, S. High-Temperature Structural Evolution of Caesium and Rubidium Triiodoplumbates. *J. Phys. Chem. Solids* **2008**, *69* (10), 2520–2526.
(15) Fadla, M. A.; Bentria, B.; Dahame, T.; Benghia, A. First-Principles Investigation on the Stability and Material Properties of All-Inorganic Cesium Lead Iodide Perovskites CsPbI3 Polymorphs. *Phys. B Condens. Matter* **2020**, *585*, 412118.
(16) Skibsted, J.; Vosegaard, T.; Bildsøe, H.; Jakobsen, H. J. 133Cs Chemical Shielding Anisotropies and Quadrupole Couplings from Magic-Angle Spinning NMR of Cesium Salts. *J. Phys. Chem.* **1996**, *100* (36), 14872–14881.
(17) Power, W. P.; Mooibroek, S.; Wasylishen, R. E.; Cameron, T. S. Cesium-133 Single-Crystal NMR Study of Cesium Chromate. *J. Phys. Chem.* **1994**, *98* (6), 1552–1560.
(18) Power, W. P.; Wasylishen, R. E.; Mooibroek, S.; Pettitt, B. A.; Danchura, W. Simulation of NMR Powder Line Shapes of Quadrupolar Nuclei with Half-Integer Spin at Low-Symmetry Sites. *J. Phys. Chem.* **1990**, *94* (2), 591–598.
(19) Kroeker, S.; Schuller, S.; Wren, J. E.; Greer, B. J.; Mesbah, A. 133 Cs and 23 Na MAS NMR Spectroscopy of Molybdate Crystallization in Model Nuclear Glasses. *J. Am. Ceram. Soc.* **2016**, *99* (5), 1557–1564.
(20) Wu, G.; Terskikh, V. A Multinuclear Solid-State NMR Study of Alkali Metal Ions in Tetraphenylborate Salts, M [BPh4](M= Na, K, Rb and Cs): What Is the NMR Signature of Cation− π Interactions? *J. Phys. Chem. A* **2008**, *112* (41), 10359–10364.
(21) Aguiar, P. M. Multinuclear Magnetic Resonance Investigations of Structure and Order in Borates and Metal Cyanides. **2007**.
(22) Czernek, J.; Brus, J. Describing the Anisotropic 133Cs Solid State NMR Interactions in Cesium Chromate. *Chem. Phys. Lett.* **2017**, *684*, 8–13.
(23) Perras, F. A.; Bryce, D. L. Multinuclear Magnetic Resonance Crystallographic Structure Refinement and Cross-Validation Using Experimental and Computed Electric Field Gradients: Application to Na2Al2B2O7. *J. Phys. Chem. C* **2012**, *116* (36), 19472–19482.
(24) Stone, N. Table of Nuclear Electric Quadrupole Moments. *At. Data Nucl. Data Tables* **2016**, *111*, 1–28.
(25) Cheng, J. T.; Edwards, J. C.; Ellis, P. D. Measurement of Quadrupolar Coupling Constants, Shielding Tensor Elements and the Relative Orientation of Quadrupolar and Shielding Tensor Principal Axis

Systems for Rubidium-87 and Rubidium-85 Nuclei in Rubidium Salts by Solid-State NMR. *J. Phys. Chem.* **1990**, *94* (2), 553–561.

(26) Vosegaard, T.; Skibsted, J.; Bildsøe, H.; Jakobsen, H. J. Quadrupole Coupling and Anisotropic Shielding from Single-Crystal NMR of the Central Transition for Quadrupolar Nuclei. 87Rb NMR of RbClO4and Rb2SO4. *J. Magn. Reson. A* **1996**, *122* (2), 111–119.

(27) Vosegaard, T.; Skibsted, J.; Bildsoe, H.; Jakobsen, H. J. Combined Effect of Second-Order Quadrupole Coupling and Chemical Shielding Anisotropy on the Central Transition in MAS NMR of Quadrupolar Nuclei. 87Rb MAS NMR of RbClO4. *J. Phys. Chem.* **1995**, *99* (27), 10731–10735.

(28) Hayashi, S.; Hayamizu, K. Accurate Determination of NMR Chemical Shifts in Alkali Halides and Their Correlation with Structural Factors. *Bull. Chem. Soc. Jpn.* **1990**, *63* (3), 913–919.

(29) Haase, A.; Kerber, M.; Kessler, D.; Kronenbitter, J.; Krüger, H.; Lutz, O.; Müller, M.; Nolle, A. Nuclear Magnetic Shielding and Quadrupole Coupling of 133Cs in Cesium Salt Powders. *Z. Für Naturforschung A* **1977**, *32* (9), 952–956.

(30) Mooibroek, S.; Wasylishen, R. E.; Dickson, R.; Facey, G.; Pettitt, B. A. Simultaneous Observation of Shielding Anisotropies and Quadrupolar Splittings in Solid State 133Cs NMR Spectra. *J. Magn. Reson. 1969* **1986**, *66* (3), 542–545.

(31) Wong, A.; Sham, S.; Wang, S.; Wu, G. A Solid-State 133Cs Nuclear Magnetic Resonance and X-Ray Crystallographic Study of Cesium Complexes with Macrocyclic Ligands. *Can. J. Chem.* **2000**, *78* (7), 975–985.

(32) Kubicki, D. J.; Prochowicz, D.; Hofstetter, A.; Zakeeruddin, S. M.; Grätzel, M.; Emsley, L. Phase Segregation in Cs-, Rb-and K-Doped Mixed-Cation (MA) x (FA) 1–x PbI3 Hybrid Perovskites from Solid-State NMR. *J. Am. Chem. Soc.* **2017**, *139* (40), 14173–14180.

(33) Haq, I. U.; Ali, A.; Khan, I. Investigation of NMR Shielding, EFG and Structural Relaxation of All-Inorganic Pb-Based Ruddlesden Popper Halide Perovskites. *Appl. Phys. A* **2023**, *129* (12), 875.